\documentclass[10pt,twocolumn,superscriptaddress,showpacs,aps,amsmath,amssymb,nofootinbib,pra]{revtex4-1}
\pdfoutput=1

\usepackage[utf8]{inputenc}

\usepackage{bm}
\usepackage{textcomp}
\usepackage{bbm}
\usepackage{enumerate}

\usepackage{tikz}
\usetikzlibrary{calc}

\usepackage{hyperref}
\hypersetup{%
pdftitle={Effects of gate errors in digital quantum simulations of fermionic systems},%
pdfauthor={Jan-Michael Reiner, Sebastian Zanker, Iris Schwenk, Juha Lepp\"akangas, Frank Wilhelm-Mauch, Gerd Sch\"on, and Michael Marthaler},%
colorlinks=true,%
linkcolor=blue,%
urlcolor=blue,%
citecolor=blue
}

\newcommand{\e}{\mathrm{e}}
\newcommand{\im}{\mathrm{i}}

\newcommand{\Ad}[2]{{\mathrm{Ad}_{#1} (#2)}}
\newcommand{\op}{A}
\newcommand{\gate}[1]{\texttt{#1}}
\renewcommand{\sp}[1]{\sigma^+_{#1}} 
\newcommand{\sm}[1]{\sigma^-_{#1}}
\newcounter{enumitemp}

\newcommand{\comment}[1]{{\color[rgb]{.8, .4, .0}\ #1}}
\renewcommand{\comment}[1]{}

\begin{document}

\title{Effects of gate errors in digital quantum simulations of fermionic systems}

\author{Jan-Michael Reiner}
\affiliation {Institut f\"ur Theoretische Festk\"orperphysik, Karlsruhe Institute of Technology (KIT), 76131 Karlsruhe, Germany}

\author{Sebastian Zanker}
\affiliation {Institut f\"ur Theoretische Festk\"orperphysik, Karlsruhe Institute of Technology (KIT), 76131 Karlsruhe, Germany}

\author{Iris Schwenk}
\affiliation {Institut f\"ur Theoretische Festk\"orperphysik, Karlsruhe Institute of Technology (KIT), 76131 Karlsruhe, Germany}

\author{Juha Lepp\"akangas}
\affiliation {Institut f\"ur Theoretische Festk\"orperphysik, Karlsruhe Institute of Technology (KIT), 76131 Karlsruhe, Germany}

\author{Frank Wilhelm-Mauch}
\affiliation {Theoretical Physics, Saarland University, 66123 Saarbr\"ucken, Germany}

\author{Gerd Sch\"on}
\affiliation {Institut f\"ur Theoretische Festk\"orperphysik, Karlsruhe Institute of Technology (KIT), 76131 Karlsruhe, Germany}
\affiliation {Institute of Nanotechnology, Karlsruhe Institute of Technology (KIT), 76021 Karlsruhe, Germany}

\author{Michael Marthaler}
\affiliation {Institut f\"ur Theoretische Festk\"orperphysik, Karlsruhe Institute of Technology (KIT), 76131 Karlsruhe, Germany}
\affiliation {Theoretical Physics, Saarland University, 66123 Saarbr\"ucken, Germany}
\affiliation {Institut f\"ur Theorie der Kondensierten Materie, Karlsruhe Institute of Technology (KIT), 76131 Karlsruhe, Germany}

\date{\today}

\begin{abstract}
Digital quantum simulations offer exciting perspectives for the study of fermionic systems such as molecules or lattice models. However, with quantum error correction still being out of reach with present-day technology, a non-vanishing error rate is inevitable. We study the influence of gate errors on simulations of the Trotterized time evolution of the quantum system with focus on the fermionic Hubbard model. Specifically, we consider the effect of stochastic over-rotations in the applied gates. Depending on the particular algorithm implemented such gate errors may lead to a time evolution that corresponds to a disordered fermionic system, or they may correspond to unphysical errors, e.g., violate particle number conservation. We substantiate our analysis by numerical simulations of model systems. In addition we establish the relation between the gate fidelity and the strength of the over-rotations in a Trotterized quantum simulation. Based on this we provide estimates for the maximum number of Trotter steps which can be performed with sufficient accuracy for a given algorithm. This in turn implies, apart from obvious limitations on the maximum time of the simulation, also limits on the system size which can be handled.
\end{abstract}


\maketitle

\section{Introduction}

Quantum devices designed to simulate quantum systems have come a long way since first ideas were formulated decades ago~\cite{feynman_simulating_1982,deutsch_quantum_1985,Cirac_Review,Quantum_Simulator_EPJ}. Impressive progress was demonstrated in recent experiments with, e.g., ultra-cold gases~\cite{Cold_Gases_Simulator,Cirac_Cold_gases,Fermi_Sea_Heidelberg}, trapped ions~\cite{350_Spin_simulator,Cirac_PRL,Quantum_Magnet_Schaetz,Ion_Simulator}, or superconducting circuits~\cite{Tian_Simu_1,Tian_Simu_2,Solan_digital_analog,Photosynthesis_Super,Fermi_Simulation_super_qubits}. While it is the goal to simulate large systems which cannot be investigated by classical means, so far most experiments are still in a proof-of-principle state. Scaling to larger simulators, i.e., a higher number of qubits, is feasible~\cite{Scalable_Devoret}. But there still remain the problems due to decoherence and errors when executing the quantum algorithms. Quantum error correction offers a route to resolve the issue of decoherence~\cite{Devitt_error_correction}, and qubits with fidelities at the threshold for the implementation of quantum error correction have been demonstrated~\cite{Marinis_Threshold,Blatt_error_correction}. But at present the number of qubits required for full quantum error correction appears prohibitive~\cite{Fowler_Surface_Code,Scalable_Hensinger}. 

In a situation where only small-size and imperfect quantum simulators are within reach, it is crucial to gain a better understanding of the effect of errors on their performance. For some examples, methods to estimate the quality of quantum simulators with errors have been suggested~\cite{Certification_Eisert,Certification_Marthaler}, and proposals for error reduction exist~\cite{Correction_IBM,Correction_Marthaler}. However, all in all it remains largely unexplored how errors will affect the results of quantum simulations, and even whether the results obtained in this way have a physical meaning~\cite{Can_We_trust_quantum_simulator}.

In this paper we evaluate the effect of gate errors on digital quantum simulations. Digital simulations are highly valued due to their wide range of applicability~\cite{Simulation_Lloyd,Simulation_Aspuru_Guzik,Simulation_Zoller}, e.g., for quantum chemistry or many-body physics. Specifically, we study simulations of fermionic systems, such as the Hubbard model, which are widely considered to be prime targets of quantum simulations~\cite{Hubbard_Frank,Hubbard_Solano}. In a digital quantum simulation the time evolution operator of the system is evaluated by splitting the Hamiltonian into parts acting in smaller subspaces via the Trotter expansion and, instead of a continuous time evolution, the simulator implements a succession of short-time Trotter steps. This short-time evolution in the appropriate subspace is then simulated by an algorithm based on a sequence of quantum gates which are available on the hardware level. The quality of the Trotter expansion requires very short time steps, and hence a very large number of gates~\cite{Simulation_Troyer_1}. It cannot be avoided that the gates are subject to some errors. Even if they are small for each individual gate, due to their large number a substantial error will accumulate, and the gain from implementing finer time steps will be limited~\cite{Errors_Munro}.

In addition, we address the question whether gate errors in a digital quantum simulation have a \emph{physical meaning}. Generally, the exact form of gate errors is not well known. Standard gates perform a rotation of the many-qubit state. We, therefore, model the gate errors as over-rotations (or under-rotations). For such gate errors we show that the simulation, rather than modeling the time evolution under the original Hamiltonian $H$, models the time evolution under an \emph{effective Hamiltonian} $H + \delta H$. The added term $\delta H$ is proportional to the strength of the over-rotations; its specific details depend on the chosen algorithm. We show that it often can be interpreted as a disorder term. However, other algorithms may introduce contributions $\delta H$ such that the physical properties of the effective Hamiltonian are very different from the original one.

The magnitude of the over-rotation can be related to the gate fidelity. Specifically, as shown in Appx.~\ref{sec:Appx-fidelity}, an over-rotation by an angle $\delta \varphi$ reduces the minimal gate fidelity $\mathcal{F}_\mathrm{min}$ (i.e., the fidelity minimized with respect to all possible input states) to the value $\mathcal{F}_\mathrm{min} = \cos(\delta \varphi)$. Reversing this relation, $|\delta\varphi| \approx \sqrt{2(1-\mathcal{F}_\mathrm{min})}$, shows that the fidelity has to be very close to $100\,\%$ to keep the gate errors small. Typically the over-rotation angles are stochastic variables. Assuming that in independent runs of the experiment the over-rotations vanish on average with a given variance $\mathrm{Var}(\delta\varphi)$, we find for the average minimal fidelity $\overline{\mathcal{F}}_\mathrm{min} = 1- \mathrm{Var}(\delta\varphi)^2/2$. It is this quantity which would be analyzed in an experiment.

We find the extra contribution to the Hamiltonian introduced by the errors scales as $|\delta H | \propto n \mathrm{Var}(\delta \varphi)$, i.e., it increases proportional to the number of Trotter steps $n$. Note that this unfortunate property is a consequence of the fact that the gate errors happen in each Trotter step and do not get weaker if one chooses shorter time steps. In addition, with increasing system size one has to perform an increasing number of gates per Trotter step. Labeling the number of gates that contribute a disorder term per Trotter step by $M$ (as we discuss below this is roughly the number of two-qubit gates per Trotter step), we arrive at the following limitation
\begin{align}
n M < \frac{1}{\sqrt{2(1-\overline{\mathcal{F}}_\mathrm{min})}}.
\end{align}
Note that this is a worst-case estimate. Depending on how errors of various gates add up, the number $M$ could effectively be much smaller. This is discussed in more detail in Sec.~\ref{sec:Fidelity-adiabatic-prep}. 

\begin{figure}
\centering
\includegraphics[width=\columnwidth]{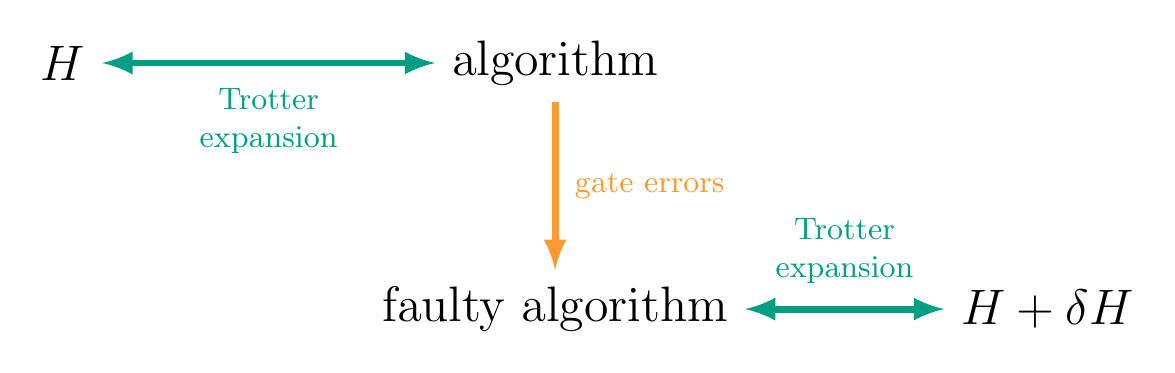}
\caption{
The structure and message of this paper: A (fermionic) system with Hamiltonian $H$ is simulated through digital quantum simulation based on Trotterization of the time evolution. Gate errors in the steps of this algorithm arise in the form of over-rotations with strength related to the fidelity of the gates. The resulting ``faulty algorithm'' is then -- within the Trotter approximation -- equivalent to the simulation of a system with Hamiltonian $H + \delta H$. We evaluate the structure of $\delta H$. It is sensitive to the specific algorithm and often describes disorder.
}
\label{fig:intro-summary}
\end{figure}

The present paper is organized as follows: In the following section we will state our model assumptions and explain how gate errors in a digital quantum simulation effectively lead to a time evolution under a modified Hamiltonian. Fig.~\ref{fig:intro-summary} gives a structural overview of our approach and findings. Thereafter we illustrate the method and results by considering a model system. We show that different choices of the algorithms lead to different effective Hamiltonians. We substantiate our findings by a numerical analysis of a minimal model system for several error strengths which we compare to present-day experimentally achieved gate fidelities. We also study the trade-off between between a finer Trotterization, i.e., less Trotter expansion error, and the resulting necessity to run more faulty gates, i.e., more impact through gate errors. Finally, we add some comments on the effect of errors in the adiabatic state preparation. We conclude with a summary and outlook.

\section{Model and method}

\subsection{Assumptions}\label{sec:Assumptions}

To be able to make quantitative predictions we have to specify the method used in the digital quantum simulation and the specific nature of the gate errors. Our primary assumptions and approximations are the following:
\begin{enumerate}[(i)]	
	\item The only type of errors in gates are over-rotations.\label{assumption-over-rotations}
	\item We consider algorithms based on the Trotter expansion. The strength (i.e., variance) of the gate errors is such that, in comparison, second order contributions in the Trotter expansion can be neglected.\label{assumption-Trotter-approximation}
	\setcounter{enumitemp}{\value{enumi}}
\end{enumerate}
Additionally we assume -- although this could be easily generalized:
\begin{enumerate}[(i)]
	\setcounter{enumi}{\value{enumitemp}}
	\item Errors in single-qubit gates can be neglected as compared to those occuring in gates between two or more qubits.\label{assumption-perfect-one-qubits-gates}
	\item The over-rotations in each gate are independent and may vary with time.\label{assumption-variation-time}
\end{enumerate}

By assumption~(\ref{assumption-over-rotations}) we imply that the needed gates are performed with the correct interaction between qubits, but there is, e.g., an uncertainty in the interaction time~\cite{willsch_gate-error_2017}. Since a gate (expressed as an exponential of Pauli operators or products thereof) can be seen as a rotation of the state of the qubit register, an error causes an over-rotation (or under-rotation). We also imply that the needed interaction is realized \emph{intrinsically} as a physical interaction on the hardware-level, or that the gate (e.g., an off-axis rotation) can to be decomposed into several intrinsic gates. In such cases the over-rotations in individual gates, in general, cannot be combined to an over-rotation in the composite one. More explicitly, we write the gate as an exponential $\e^{\mathrm{i} \varphi \op}$, with an dimensionless operator $\op$ with unit norm, $||\op|| = 1$ where $||\cdot||$ denotes the standard induced operator norm,\footnote{The normalization does not fix the angle of rotation but imposes an upper bound to it. Consider Pauli operators $\sigma^+ \sigma^-$ and $\sigma^z$: Even though $||\sigma^+ \sigma^-|| = ||\sigma^z|| = 1$, the first operator rotates half as much as the second, since $\sigma^+ \sigma^- = \frac{\sigma^z}{2} + \frac{1}{2}$. Below we ignore such constant factors (and global phases). A more thorough definition of $A$ is given in Appx.~\ref{sec:Appx-fidelity}.} characterizing the type of interaction of the gate, and an angle $\varphi$. For example, for an \gate{iSWAP} gate between qubits $j$ and $k$, one has $\op = \sigma^+_j \sigma^-_k + \sigma^+_k \sigma^-_j$ (where $\sigma^\pm_{j,k}$ are the ladder operators of the qubits $j,k$) and $\varphi = \pi/2$. An over-rotation is characterized by an addition $\delta \varphi$ to the angle, such that the faulty gate is $\e^{\mathrm{i} (\varphi + \delta \varphi) \op}$.

Assumption~(\ref{assumption-Trotter-approximation}) implies bounds for $\delta \varphi$. We restrict ourselves to algorithms simulating the time evolution under a Hamiltonian $H$, or in general $H(t)$, making use of the Trotter expansion. It assumes that the time evolution operator during time $0\le t\le \tau$ is broken into a product of a large number of $n$ short segments of length $\tau/n$ (which is a way to assure the proper time order in the time-evolution operator). It further exploits that the usually complicated many-body Hamiltonian $H$ can be decomposed into $N$ parts, $H = \sum_{j=1}^N H_j$, each acting in a much reduced Hilbert space. If the Hamiltonian does not depend on time we have
\begin{align}
\e^{-\mathrm{i} H \tau} = \e^{-\mathrm{i} \sum_{j=1}^N H_j \tau} &= \left(\prod_{j=1}^N \e^{-\mathrm{i} H_j \frac{\tau}{n}} + \mathcal{O}\Big(\big(\frac{g\tau}{n}\big)^2\Big) \right)^n\nonumber \\
&= \left(\prod_{j=1}^N \e^{-\mathrm{i} H_j \frac{\tau}{n}}\right)^n + \mathcal{O}\Big(\frac{(g\tau)^2}{n}\Big),\label{eq:Trotter-expansion}
\end{align}
The generalization to a time-dependent Hamiltonian is obvious. Above we introduced $g=\max_j ||H_j||$, the largest energy scale of the Hamiltonian. If the number of Trotter steps is chosen sufficiently large, such that $\frac{g\tau}{n} \ll 1$, the first order Trotter expansion is sufficient. If each $H_j$ can be implemented and controlled individually by the hardware of the quantum simulator, one is able to simulate the time evolution up to the time $\tau$ in this way.
 
Rewriting the exponents in Eq.~\eqref{eq:Trotter-expansion} as $-\mathrm{i} H_j \frac{\tau}{n} = \mathrm{i} \frac{-g\tau}{n} \frac{H_j}{g}$ and noting that $H_j/g$ has (less than) unit norm, shows that -- in analogy to what we discussed under assumption~(\ref{assumption-over-rotations}) -- the Trotterization introduces angles of order $\frac{g\tau}{n}$. The over-rotation $\delta \varphi$ should be of smaller magnitude
\begin{align}
|\delta \varphi| \le \frac{g\tau}{n}.
\label{eq:limit on deltaphi}
\end{align}
In this case, order $\mathcal{O} (\delta \varphi^2)$ terms and products between $\delta \varphi$ and $\frac{g\tau}{n}$ can be neglected inside a Trotter step, consistent with the Trotter approximation.

The optional assumption~(\ref{assumption-perfect-one-qubits-gates}) is motivated by the observation that single-qubit gates usually have a significantly better fidelity than two-qubit gates, which are more difficult to optimize~\cite{Marinis_Threshold,ballance_high-fidelity_2016,rong_experimental_2015,gaebler_high-fidelity_2016,gefen_enhancing_2017}. Furthermore, we note that, e.g., the Hubbard model with equal on-site energies can be simulated using only two-qubit gates. Single-qubit gates are needed if only a limited set of two-qubit gates are available. For example, if the hardware allows for $XX$ interaction between qubits, on can, with additional single-qubit rotations, also implement a $ZZ$ interaction. Such $XX$ interactions arise, e.g., between nearest-neighbor capacitively coupled transmon qubits~\cite{koch_charge-insensitive_2007,Tian_Simu_2}, or similarly in trapped-ion architectures~\cite{debnath_demonstration_2016}. All conclusion drawn in the following would not change if an assumed interaction would involve additional (error-free) single-qubit gates.

As expressed by assumption~(\ref{assumption-variation-time}) we allow the gate errors to vary with time.
In the numerical simulation performed below we assume that the errors are normal-distributed with zero mean. Particularly, we will investigate the effect of errors which are averaged over different runs of an algorithm, each one performed with a different realization of the (random) errors. However, we should note that at no point in our discussion this is strictly necessary. A constant error will lead to a constant perturbative correction to the Hamiltonian, which can be estimated. On top of it, it should be possible to reduce the effect of constant errors more efficiently.

\subsection{Method}\label{sec:Method}

With the assumptions specified, we are now ready to analyze the effects of gate errors due to over-rotations. There are two distinct cases:

\subparagraph{Case 1:} Consider a Hamiltonian $H = H_1 + H_2$, where $H_{(1,2)} = g_{(1,2)} A_{(1,2)}$ with $||A_{(1,2)}|| = 1$. For simplicity of the notation in the following we assume $g_1 = g_2 = g$. Also note that the following discussion is easily generalized to a situation with more terms. In this first, simple case, we further assume that the quantum simulator allows performing directly -- with the available hardware by single gates -- the Trotter steps arising from the interactions $A_{(1,2)}$. In this case, the simulation of the time evolution under $H$ is based on the Trotter expansion,
\begin{align}
\e^{-\mathrm{i} H \tau} = \bigg( \e^{-\mathrm{i} \frac{g\tau}{n} A_1} \e^{-\mathrm{i} \frac{g\tau}{n} A_2} + \mathcal{O}\Big(\big(\frac{g\tau}{n}\big)^2\Big) \bigg)^n \, .
\end{align}
Gate errors due to independent over-rotations $\delta \varphi_{(1,2)m}$ during the $m^\mathrm{th}$ Trotter step lead to the modification
\begin{align}
& \e^{-\mathrm{i} \frac{g\tau}{n} A_1} \e^{-\mathrm{i} \frac{g\tau}{n} A_2} \nonumber \\
\mapsto{}& \e^{-\mathrm{i} (\frac{g\tau}{n} + \delta \varphi_{1m}) A_1} \e^{-\mathrm{i} (\frac{g\tau}{n} + \delta \varphi_{2m}) A_2} \nonumber \\
={}& \e^{-\mathrm{i} \frac{g\tau}{n} A_1} \e^{-\mathrm{i} \frac{g\tau}{n} A_2} \e^{-\mathrm{i} \delta \varphi_{1m} A_1} \e^{-\mathrm{i} \delta \varphi_{2m} A_2} + \mathcal{O} \Big(\big(\frac{g\tau}{n}\big)^2\Big).
\end{align}
In the last step we made use of Eq.~\eqref{eq:limit on deltaphi}.
Comparing to the time-dependent generalization of Eq.~\eqref{eq:Trotter-expansion}, we find that the over-rotations lead to the simulation of the time evolution under an effective Hamiltonian $H + \delta H(t)$, where
\begin{align}\label{eq:disorder-terms-1}
\delta H (t) = \frac{n}{\tau} \delta \varphi_1 (t) A_1 + \frac{n}{\tau} \delta \varphi_2 (t) A_2 \, .
\end{align}
with $\delta \varphi_{(1,2)} (t) = \delta \varphi_{(1,2)m}$ for $t \in [\frac{\tau}{n} (m-1), \frac{\tau}{n}m)$. We note that the contributions from the errors scale linearly with $n$, i.e., they become larger when the Trotterization is based on more, finer steps. Therefore, as noted in Ref.~\cite{Errors_Munro}, the gain from a finer Trotterization is limited. On the other hand, we note that the corrections are still small compared to the largest energy scale of the original Hamiltonian, $\frac{n}{\tau} |\delta \varphi_{(1,2)}(t)|\leq g$.

\subparagraph{Case 2:} The situation is more complex if an interaction, e.g., $A_1$, cannot be implemented directly by gates available on the hardware-level of the quantum simulator and a decomposition involving additional two-qubit gates is required. They are also subject to errors due to over-rotations. An example could be a chain of qubits with only nearest-neighbor $XX$-couplings. To simulate a Hamiltonian with $XX$ interactions between next-nearest neighbors one needs to introduce, e.g., swap gates before and after the available nearest-neighbor $XX$ interaction. Other examples for this scenario are simulations which require gates such as \gate{CNOT}s, \gate{iSWAP}s, etc. 

Let us assume that a Trotter step with interaction $A_1$ is decomposed as 
\begin{align}\label{eq:gate-decomposition}
\e^{-\im \frac{g\tau}{n} A_1} = \e^{\im \varphi C} \e^{-\im \frac{g\tau}{n} B_1} \e^{\im \varphi' C'},
\end{align}
where $B_1$ and $C^{(\prime)}$ can be implemented on the hardware level. Again we choose $||B_1||=||C^{(\prime)}||=1$, and the angles $|\varphi^{(\prime)}|$ are typically of order $\mathcal O (1)$, i.e., much larger than $\frac{g\tau}{n}$. It is more difficult to treat gate errors now. Commutators between exponentials with exponents of order $\mathcal O (\delta \varphi)$ of the gate errors and exponential with exponents of order $\mathcal{O} (1)$ have to be taken into account.

To calculate these commutators, we use the relation
\begin{align}
\e^X \e^Y \e^{-X} = \e^{\e^X Y \e^{-X}} = \e^\Ad{\e^X}{Y},
\end{align}
where for later convenience we introduced the notation $\Ad{\e^X}{Y} = \e^X Y \e^{-X}$ for the adjoint representation. It follows that
\begin{align}\label{eq:adjoint-commutation}
\e^X \e^Y &= \e^\Ad{\e^X}{Y} \e^X,\nonumber\\
\e^Y \e^X &= \e^X \e^\Ad{\e^{-X}}{Y} \, .
\end{align}
Errors due to over-rotations, $\delta\varphi, \delta\varphi_1, \delta\varphi'$, in the gate~\eqref{eq:gate-decomposition} will enter in each of the three exponents. To evaluate their effects we use the above relations to commute the errors out of the original gate decomposition and arrive at
\begin{align}\label{eq:case-2-commutation}
&\e^{\im \varphi C} \e^{-\im \frac{g\tau}{n} B_1} \e^{\im \varphi' C'} \nonumber \\
\mapsto{}& \e^{\im (\varphi + \delta \varphi) C} \e^{-\im (\frac{g\tau}{n} + \delta \varphi_1) B_1} \e^{\im (\varphi' + \delta \varphi') C'} \nonumber \\
={}& \e^{\im \delta \varphi C} \e^{\im \varphi C} \e^{-\im \delta \varphi_1 B_1} \e^{-\im \frac{g\tau}{n} B_1} \e^{\im \varphi' C'} \e^{\im \delta \varphi' C'} \nonumber \\
={}& \e^{\im \delta \varphi C} \e^{\Ad{\e^{\im \varphi C}}{-\im \delta \varphi_1 B_1}} \underbrace{\e^{\im \varphi C} \e^{-\im \frac{g\tau}{n} B_1} \e^{\im \varphi' C'}}_{=\e^{-\im \frac{g\tau}{n} A_1}} \e^{\im \delta \varphi' C'} \nonumber \\
={}& \e^{-\im \frac{g\tau}{n} A_1} \e^{\im \delta \varphi C} \e^{\im \delta \varphi' C'} \e^{-\im \delta \varphi_1 \Ad{\e^{\im \varphi C}}{B_1}} + \mathcal{O} \left(\big(\frac{g\tau}{n}\big)^2\right).
\end{align}
All the exponents are now of order $\mathcal O (\frac{g\tau}{n})$ such that commuting of the exponentials contributes negligible errors of order $\mathcal{O}((\frac{g\tau}{n})^2)$. Hence, we can employ the Trotter expansion~\eqref{eq:Trotter-expansion} and find the effective time-dependent Hamiltonian $H + \delta H (t)$ that is simulated, with
\begin{align}\label{eq:disorder-terms-2}
\delta H(t) = - \frac{n}{\tau} \delta \varphi (t) C - \frac{n}{\tau} \delta \varphi'(t) C' + \frac{n}{\tau} \delta \varphi_1(t) \Ad{\e^{\im \varphi C}}{B_1}.
\end{align}
The presented scheme can be applied iteratively for higher stage gate decompositions. Therefore, it is a scalable approach where algorithms can be analyzed piecewise.

The non-trivial addition as compared to Case~1 is the term $\Ad{\e^{\im \varphi C}}{B_1}= \e^{\im \varphi C} B_1 \e^{-\im \varphi C}$. One may evaluate this by using Hadamard's lemma (related to the Baker-Campbell-Hausdorff formula):
\begin{align}
&\e^{X} Y \e^{-X} \nonumber \\
={}& Y + [X, Y] + \frac{1}{2} [X, [X, Y]] + \frac{1}{3!} [X, [X, [X, Y]]] + \dots
\end{align}
However, since in the Trotter expansion we commute quantum gates, the exponentials often have simple representations in terms of Pauli matrices. Hence, the product $\e^{\im \varphi C} B_1 \e^{-\im \varphi C}$ can be calculated by expressing $\e^{\pm \im \varphi C}$ and $B_1$ in terms of Pauli matrices. This is evident for $B_1$, since it describes an interaction between qubits, but also for $\e^{\im \varphi C}$, which might be a quantum gate like \gate{CNOT}, \gate{(i)SWAP}, \gate{CZ}, and so forth.

We have now established our method to treat gate errors in a digital quantum simulation. In order to render the description less abstract we will apply it in the next section to an example algorithm. We will explicitly show the structure of $\delta H$ that is generated by the different parts of the algorithm.

\section{Fermi-Hubbard model}
\label{sec:Example-algorithm}

\subsection{Fermi-Hubbard model with spin-flip interaction}

For illustration we will investigate now in detail the Fermi-Hubbard model for a small system consisting only of two degenerate sites including spin-flip interaction. This minimal model already allows us to illustrate the important issues. The Hamiltonian is 
\begin{align}\label{eq:Hamiltonian-fermions}
H &= U \sum_{j=1}^2 c^\dagger_{j\uparrow} c^{\phantom \dagger}_{j\uparrow} c^\dagger_{j\downarrow} c^{\phantom \dagger}_{j\downarrow} - t_1 \sum_{s=\uparrow,\downarrow} (c^\dagger_{1s} c^{\phantom \dagger}_{2s} + c^\dagger_{2s} c^{\phantom \dagger}_{1s}) \nonumber \\
&\phantom = - t_2 \sum_{j=1}^2 (c^\dagger_{j\uparrow} c^{\phantom \dagger}_{j\downarrow} + c^\dagger_{j\downarrow} c^{\phantom \dagger}_{j\uparrow}) \, ,
\end{align}
where $c^{(\dagger)}_{j,s}$ stands for the fermionic annihilation (creation) operator of site $j$ with spin $s$. The on-site interaction energy is $U$, the hopping element between the sites is $t_1$, and spin-flips on each site occur with amplitude $t_2$. We map the fermionic system on a system of spins/qubits via the Wigner-Jordan transformation. Because of the spin-flip interaction in addition to the hopping this mapping is non-trivial~\cite{Tian_Simu_2} even for the small (one-dimensional) system considered.

To proceed we first relabel the fermionic operators,
\begin{align}
c_1 = c_{1\uparrow}, & & c_2 = c_{2\uparrow}, & & c_3 = c_{2\downarrow}, & & c_4 = c_{1\downarrow},
\end{align}
and map them on Pauli matrices $\sigma^{x,y,z}$ and the ladder operators $\sigma^\pm = \frac{1}{2} (\sigma^x \pm \mathrm{i} \sigma^y)$ via the Jordan-Wigner transformation
\begin{align}
c_j = \prod_{k=1}^{j-1} (-\sigma^z_k) \sigma^-_j \, .
\end{align}
The transformation introduces a product of $\sigma^z$ operators, which we call in the following Jordan-Wigner string. The Hamiltonian thus becomes
\begin{align}\label{eq:Hamiltonian-qubits}
H &= U (\sp 1 \sm 1 \sp 4 \sm 4 + \sp 2 \sm 2 \sp 3 \sm 3) \nonumber \\
&\phantom = - t_1 \big( (\sp 1 \sm 2 + \sm 1 \sp 2) + (\sp 3 \sm 4 + \sm 3 \sp 4) \big) \nonumber \\
&\phantom = - t_2 \big( (\sp 2 \sm 3 + \sm 2 \sp 3) + (\sp 1 \sigma^z_2 \sigma^z_3 \sm 4 + \sm 1 \sigma^z_2 \sigma^z_3 \sp 4) \big).
\end{align}

We consider the situation where the hardware allows for $XX$ and $ZZ$ interactions between each pair of qubits (note that the two are related to each other via single-qubit gates, which by assumption are free of errors). In this case every term in $H$ can be realized by a two-qubit gate, except for the non-local terms $-t_2\sigma^\pm_1 \sigma^z_2 \sigma^z_3 \sigma^\mp_4$, which need to be decomposed into multiple two-qubit gates.

\subsection{Algorithm and effects of gate errors}
\label{sec:Algorithm}

\begin{figure}
\centering
\includegraphics[width=\columnwidth]{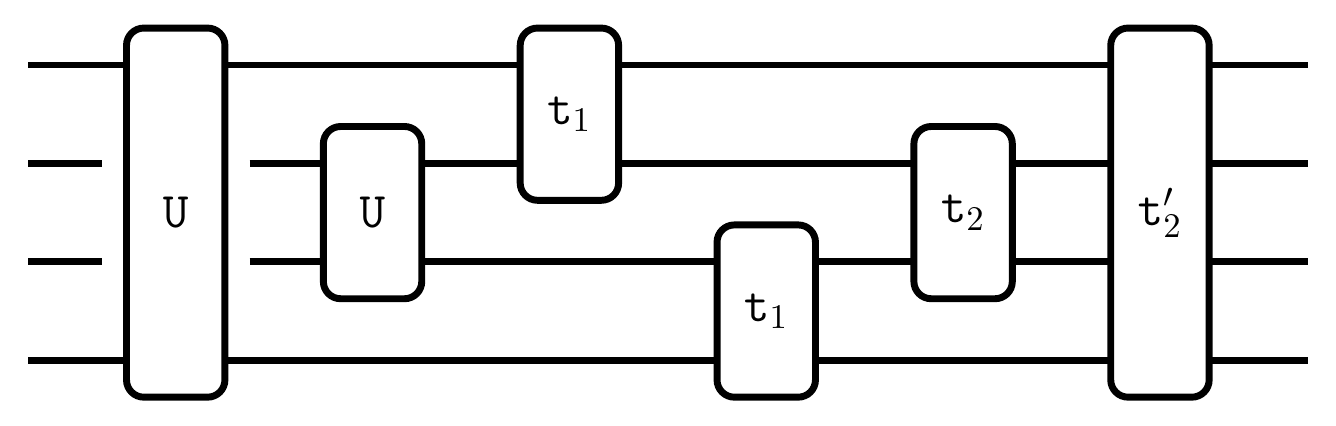}
\caption{%
One Trotter step of an algorithm simulating the time evolution under the Hamiltonian~\eqref{eq:Hamiltonian-qubits}. The horizontal lines represent the qubits and the boxes the gates of the algorithm. A qubit line connected to a gate is an in-/output of that gate. If a line is interrupted at a gate the corresponding qubit is not influenced by that gate. The two-qubit $\gate U$ gates implement the on-site interaction, the $\gate t_1$ gates the hopping terms, the $\gate t_2$ gate a part of the spin-flip interactions. The four-qubit gate $\gate t_2'$ accounts for the non-local term which arises due to the Jordan-Wigner transformation.
}
\label{fig:full-step}
\end{figure}

We implement the time evolution of the Hamiltonian~\eqref{eq:Hamiltonian-qubits} by a Trotter expansion, where the gates needed for one Trotter step are depicted in Fig.~\ref{fig:full-step}. It involves two-qubit gates labeled $\gate U$, which account for the on-site interaction, two-qubit gates labeled $\gate t_1$ for the hopping terms, and a two-qubit gate $\gate t_2$ which implements the first part of the spin-flip interaction. The non-local terms of the spin-flip interaction introduced by the Jordan-Wigner string are represented by a four-qubit gate $\gate t'_2$. We now will use the method laid out in Sec.~\ref{sec:Method} to analyze the effects of gate errors in the various parts of this algorithm and also propose an explicit decomposition of the four-qubit gate.

\subparagraph{\gate U gates.}

The first $\gate U$ gate in Fig.~\ref{fig:full-step} has the representation
\begin{align}
\gate U_{1,4} = \e^{-\im U \frac{\tau}{n} \sp 1 \sm 1 \sp 4 \sm 4},
\end{align}
where we introduced indices indicating that the interaction occurs between qubit one and four. We allow for over-rotations,
\begin{align}
\gate U_{1,4} \mapsto{}& e^{-\im (U \frac{\tau}{n} + \delta \varphi^U_{1,4}) \sp 1 \sm 1 \sp 4 \sm 4} \nonumber \\
={}& e^{-\im (U \frac{\tau}{n} + \delta \varphi^U_{1,4}) c^\dagger_1 c^{\phantom \dagger}_1 c^\dagger_4 c^{\phantom \dagger}_4}.
\end{align}
This scenario is covered by Case~1 of Sec.~\ref{sec:Method}, and we easily find the contributions to the Hamiltonian that will actually be simulated in the presence of over-rotation,
\begin{align}
H \mapsto{}& H + \frac{n}{\tau} \delta \varphi^U_{1,4}(t) \sp 1 \sm 1 \sp 4 \sm 4 \nonumber \\
={}& H + \frac{n}{\tau} \delta \varphi^U_{1,4}(t) c^\dagger_1 c^{\phantom \dagger}_1 c^\dagger_4 c^{\phantom \dagger}_4,
\end{align}
and likewise for the second gate $\gate U_{2, 3}$.

\subparagraph{$\gate t_1$ and $\gate t_2$ gates.}

The $\gate t_1$ and $\gate t_2$ gates of the algorithm in Fig.~\ref{fig:full-step} apply similar transversal interactions. Analogously to the case of \gate U gates, we label them $\gate t_{1,2}$, $\gate t_{3,4}$, and $\gate t_{2,3}$, where the first two have amplitude $t_1$ while the third has amplitude $t_2$. For instance we have
\begin{align}
\gate t_{1,2} = \e^{\im t_1 \frac{\tau}{n} (\sp 1 \sm 2 + \sm 1 \sp 2)}.
\end{align}
Again we are in the situation of Case~1 of Sec.~\ref{sec:Method}, i.e., gate errors lead to contributions of the form
\begin{align}
H \mapsto{}& H - \frac{n}{\tau} \delta\varphi^{t_1}_{1,2} (t) (\sp 1 \sm 2 + \sm 1 \sp 2) \nonumber \\
={}& H - \frac{n}{\tau} \delta\varphi^{t_1}_{1,2} (t) (c^\dagger_1 c^{\phantom \dagger}_2 + c^\dagger_2 c^{\phantom \dagger}_1),
\end{align}
and likewise for the gates $\gate t_{3,4}$ and $\gate t_{2,3}$.

\subparagraph{$\gate t'_2$ gate.}

\begin{figure}
\centering
\includegraphics[width=\columnwidth]{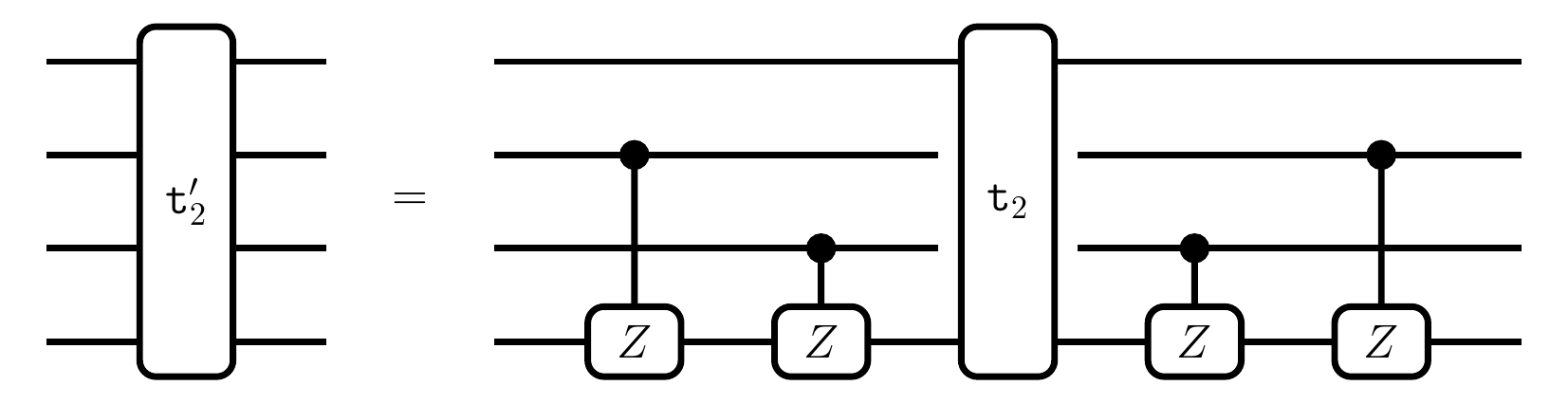}
\caption{%
The four-qubit gate $\gate t_2'$ can be decomposed into a two-qubit $\gate t_2$ gate and additional \gate{CZ}s accounting for the Jordan-Wigner string.
}
\label{fig:t2prime-subst-JW}
\end{figure}

The $\gate t'_2$ gate falls into the scope of Case~2 of Sec.~\ref{sec:Method}. We assume that the hardware allows performing two-qubit $ZZ$ and $XX$ gates between the qubits. But, the four-qubit gate $\gate t'_2$ has to be decomposed. One possible decomposition is shown in Fig.~\ref{fig:t2prime-subst-JW}. A $\gate t_2$ gate is introduced between qubits one and four, and the Jordan-Wigner strings are implemented with strings of controlled $Z$ gates.

The controlled $Z$ gates can be treated easily even in the presence of gate errors. It holds that a \gate{CZ} gate between the control qubit $j$ and the target qubit $k$ has the form $\gate{CZ}_{j,k} = \e^{\im \pi \sp j \sm j \sm k \sp k}$. The \gate{CZ} gates are diagonal and commute among each other. Hence, over-rotations are trivially shifted to the left and right of the original gate composition and contribute to the simulated Hamiltonian with terms of the form
\begin{align}
H \mapsto{}& H - \frac{n}{\tau} \delta \varphi_{j,k} (t) \sp j \sm j \sm k \sp k \nonumber \\
={}& H - \frac{n}{\tau} \delta \varphi_{j,k} (t) c^\dagger_j c^{\phantom \dagger}_j c^{\phantom \dagger}_k c^\dagger_k.
\end{align}

\begin{figure}
\centering
\includegraphics[width=\columnwidth]{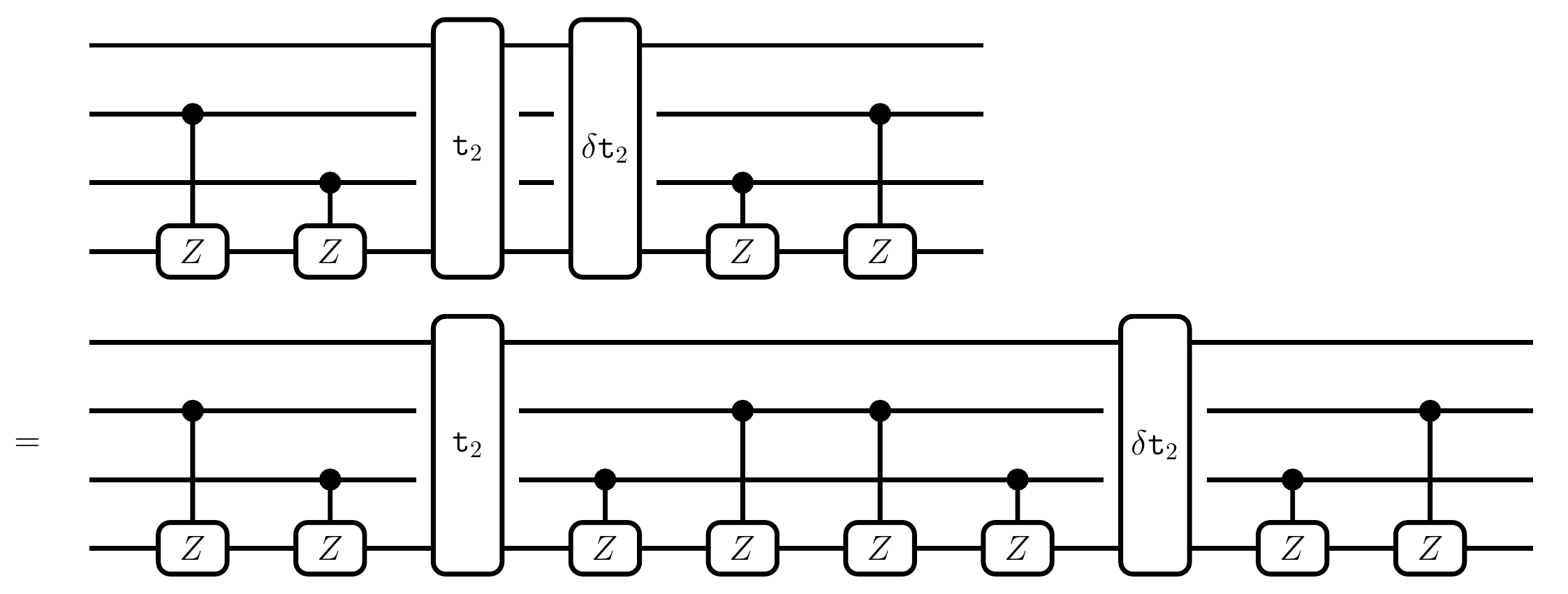}
\caption{%
The gate $\gate t_2$ gate and the error $\delta \gate t_2$ are analyzed by writing the identity as a string of virtual \gate{CZ} gates..
}
\label{fig:t2prime-subst-CZidentity}
\end{figure}

Errors in the $\gate t_2$ interaction can be analyzed by using Eq.~\eqref{eq:case-2-commutation} and evaluating the adjoint. However, in the present case there exists an easier approach: We label the gate errors $\delta \gate t_2$ and use the fact that $\gate{CZ}_{j_k}^2 = \mathbbm{1}$ to introduce additional (virtual) gates. We can see in Fig.~\ref{fig:t2prime-subst-CZidentity} how this implements Jordan-Wigner strings for $\delta \gate t_2$ such that it transforms to a gate $\delta \gate t_2'$ analogously to the transformation $\gate t_2 \mapsto \gate t_2'$. Hence, the Hamiltonian effectively simulated is
\begin{align}
H \mapsto{}& H - \frac{n}{\tau} \delta\varphi^{t_2}_{1,4} (t) (\sp 1 \sigma^z_2 \sigma^z_3 \sm 4 + \sm 1 \sigma^z_2 \sigma^z_3 \sp 4) \nonumber \\
={}& H - \frac{n}{\tau} \delta\varphi^{t_2}_{1,4} (t) (c^\dagger_1 c^{\phantom \dagger}_4 + c^\dagger_4 c^{\phantom \dagger}_1).
\end{align}

Adding up the various contributions discussed in this subsection, we find that due to the effect of over-rotations we effectively simulate a \emph{disordered} Hamiltonian. The structure and strength of the over-rotations of each gate in each Trotter step can be determined experimentally by examining each gate individually beforehand. With this knowledge one can interpret the results of the simulation such as, e.g., the spectral resolution.

\subsection{Scaling up to a larger system}
\label{sec:Scaling-up}

In this subsection, we comment on an issue that emerges when one tries to scale up the presented model system. A potential drawback of the implementation of Jordan-Wigner strings involving \gate{CZ} gates (see Fig.~\ref{fig:t2prime-subst-JW}) are interactions between qubits that are far apart in the circuit. In devices where, e.g., qubits are arranged in chains with nearest-neighbor interaction only, this becomes tedious.

\subparagraph{Jordan-Wigner strings using \gate{CNOT}s.}

\begin{figure}
\centering
\includegraphics[width=\columnwidth]{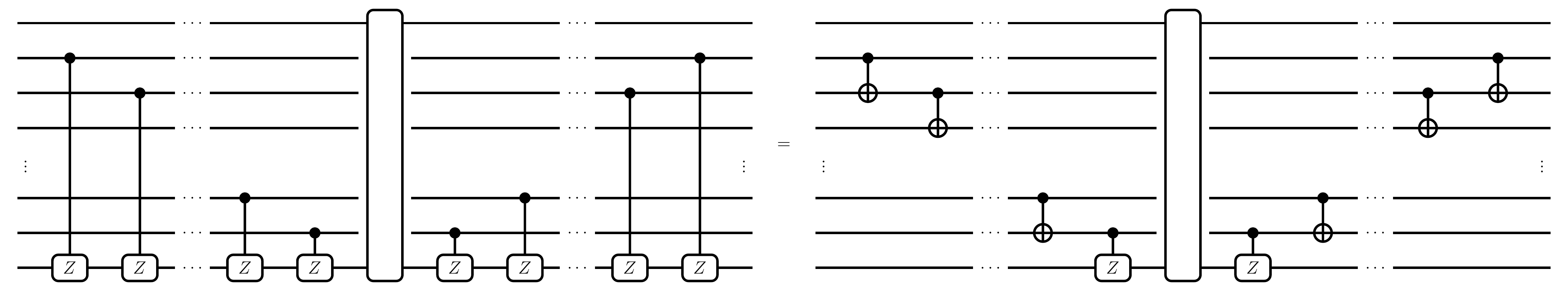}
\caption{%
Instead of using \gate{CZ} chains between far separated qubits for implementing the Jordan-Wigner string, one can use chains of \gate{CNOT}s between nearest neighbors.
}
\label{fig:CNOTs-instead-of-CZs}
\end{figure}

\begin{figure}
\centering
\includegraphics[width=.5\columnwidth]{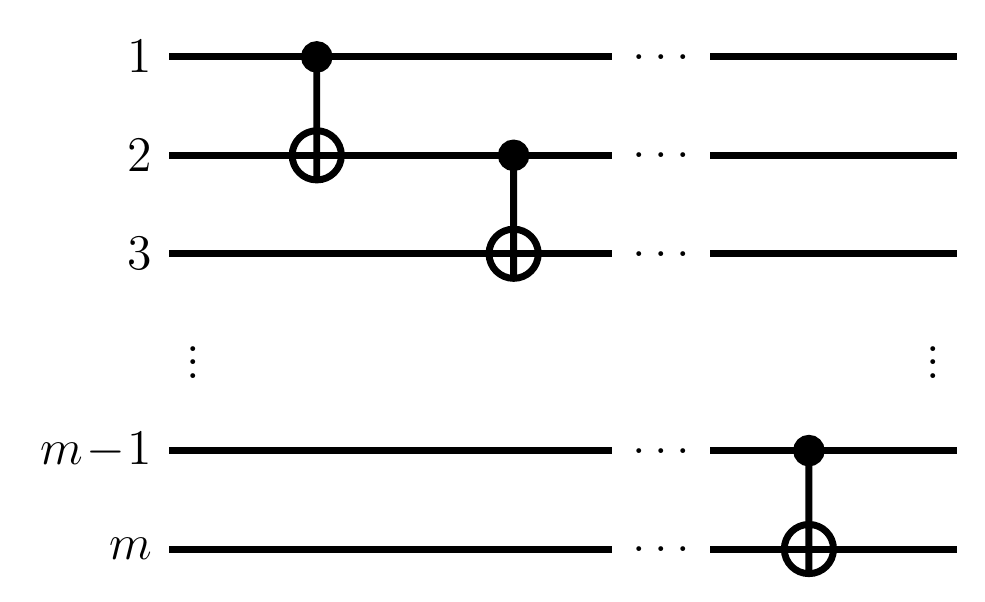}
\caption{%
A chain of \gate{CNOT} gates in an array of $m$ qubits, as it appears in Fig.~\ref{fig:CNOTs-instead-of-CZs}.
}
\label{fig:CNOT-chain}
\end{figure}

A possible solution is to employ instead of \gate{CZ}s the gates \gate{CNOT}, as depicted in Fig.~\ref{fig:CNOTs-instead-of-CZs}. Unfortunately, as we will demonstrate this has also specific disadvantages. For example, in a circuit with $ZZ$ and $XX$ interaction the \gate{CNOT} requires an $XZ$-type mixture of these interactions. This can be achieved by adding extra single-qubit gates, which are assumed to have a negligible error. However, unlike \gate{CZ} gates, the \gate{CNOT} gates do not commute among each other. A \gate{CNOT} with control and target qubit $j$ and $k$ can be written in the form
\begin{align}
\gate{CNOT}_{j,k} = \sp j \sm j \sigma^x_k + \sm j \sp j = \e^{\im \frac{\pi}{2} \sp j \sm j (1 - \sigma^x_k)}.
\end{align}
Consider a chain of \gate{CNOT}s in an array of $m$ qubits as depicted in Fig.~\ref{fig:CNOT-chain}. We allow for an over-rotation $\delta \varphi$ in the rightmost gate, which we would like to commute to the far left. This requires an analysis as discussed in Case 2 of Sec.~\ref{sec:Method}. We commute the error gate by gate using Eq.~\eqref{eq:adjoint-commutation}. The first commutation yields the exponent
\begin{align}
&\Ad{\gate{CNOT}_{m-2,m-1}}{\im \delta \varphi \sp{m-1} \sm{m-1} (1-\sigma^x_m)} \nonumber \\
={}& \im \delta \varphi (1-\sigma^x_m) \Ad{\gate{CNOT}_{m-2,m-1}}{\sp{m-1} \sm{m-1}} \nonumber \\
={}& \im \delta \varphi (1-\sigma^x_m) (\sp{m-1} \sm{m-1} - \sp{m-2} \sm{m-2})^2 \, .
\end{align}
Commuting this result one gate further gives a term
\begin{align}
{}&\Ad{\gate{CNOT}_{m-3,m-2}}{(\sp{m-1} \sm{m-1} - \sp{m-2} \sm{m-2})^2} \nonumber \\
={}& (\sp{m-1} \sm{m-1} - \Ad{\gate{CNOT}_{m-3,m-2}}{\sp{m-2} \sm{m-2}} )^2 \nonumber \\
={}& \big(\sp{m-1} \sm{m-1} - (\sp{m-2} \sm{m-2} - \sp{m-3} \sm{m-3})^2 \big)^2.
\end{align}
We can iteratively continue this process until we commuted the error to the far left. In a time evolution algorithm we see that this error contributes on the Hamiltonian level as
\begin{align}\label{eq:CNOT-chain-commutations}
\delta H (t) &= \frac{n}{\tau} \delta \varphi(t) (1-\sigma^x_m) \Sigma^\pm_{m-1},
\end{align}
with the nested operator $\Sigma^\pm_{m-1}$ defined via $\Sigma^\pm_{1} = \sp 1 \sm 1$ and $\Sigma^\pm_{j} = (\sp{j} \sm{j} - \Sigma^\pm_{j-1})^2$ (for integers $j \geq 2$).

Although it turned out to be more tedious than with the use of \gate{CZ} chains, we succeeded to analyze over-rotations in Jordan-Wigner strings by using \gate{CNOT} gates, and we could write down the spin Hamiltonian that is simulated in the presence of gate errors. However, it is now problematic to transform back into a fermionic representation. For this purpose we can apply the inverse Jordan-Wigner transformation $\sigma^+_m = \prod_{k=1}^{m-1} (2 c^\dagger_k c^{\phantom{\dagger}}_k - 1) c_m$ to Eq.~\eqref{eq:CNOT-chain-commutations} using $\sigma^x_m = \sigma^+_m + \sigma^-_m$ and $\sp j \sm j = c^\dagger_j c^{\phantom \dagger}_j$. However, we will end up with a product of many fermionic operators in the effective Hamiltonian, which raises the question about the meaning of these terms.

Up to this point, gate errors only introduced disorder terms that are of the same form as terms which were already contained in the Hamiltonian. Hence, the physical meaning was clear; we were effectively simulating a disordered system. In the case of \gate{CNOT} gates, errors take us out of the scope of either hoppings $t_{jk} c^\dagger_j c^{\phantom \dagger}_k$ (including spin flips) or interactions $V_{jklm} c^\dagger_j c^\dagger_k c^{\phantom \dagger}_l c^{\phantom \dagger}_m$. Instead, we introduce effective interactions between many sites. Specifically, these terms contain an \emph{uneven} number of fermionic operators and -- unlike the original Hubbard model -- violate particle number conservation.

Because of this problem we suggest an alternative solution to the modeling of \gate{CZ} gates between far distant qubits.

\subparagraph{\gate{iSWAP} chains.}

\begin{figure}
\centering
\includegraphics[width=\columnwidth]{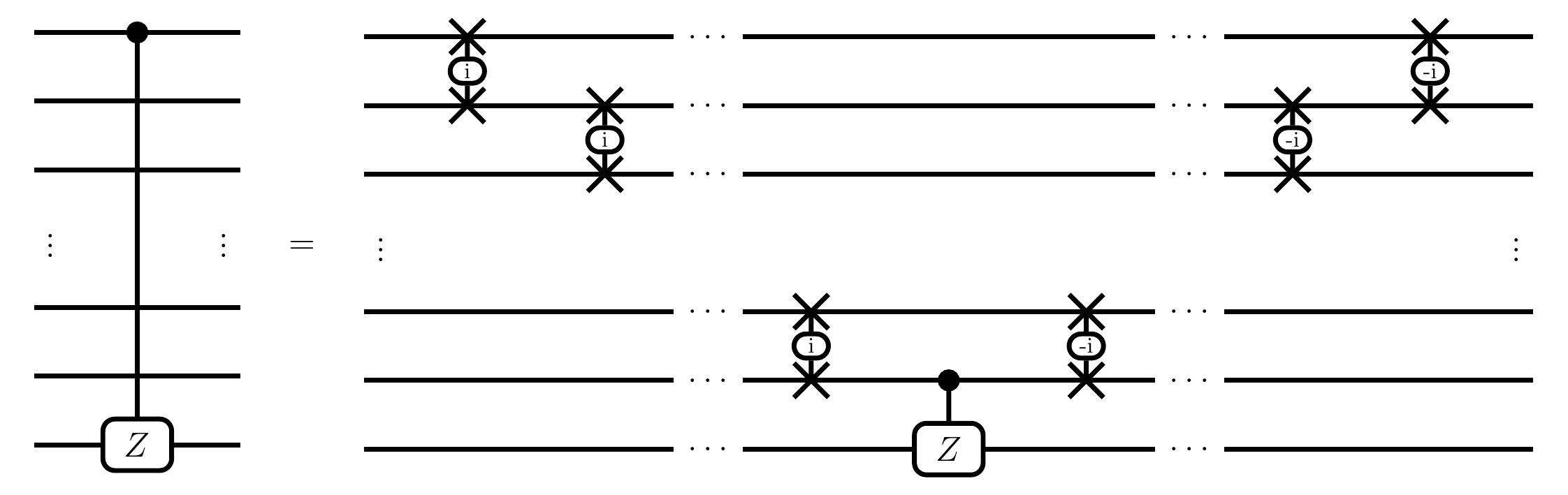}
\caption{%
If the interaction of a \gate{CZ} cannot be applied directly between two qubits that are far apart on the hardware, one can use chains of \gate{$\pm$iSWAP}s to circumvent the problem.
}
\label{fig:long-range-CZ-with-iSWAPs}
\end{figure}

An alternative approach to deal with a controlled $Z$ where control and target are far apart is by inserting $\pm \gate{iSWAP}$s as displayed in Fig.~\ref{fig:long-range-CZ-with-iSWAPs}. Doing so will not alter the form of the contributions to the Hamiltonian that stem from errors in the \gate{CZ} gates which are discussed above. This can be seen by applying additional \gate{$\pm$iSWAP}s (similar to Fig.~\ref{fig:t2prime-subst-CZidentity}).

\begin{figure}
\centering
\includegraphics[width=.5\columnwidth]{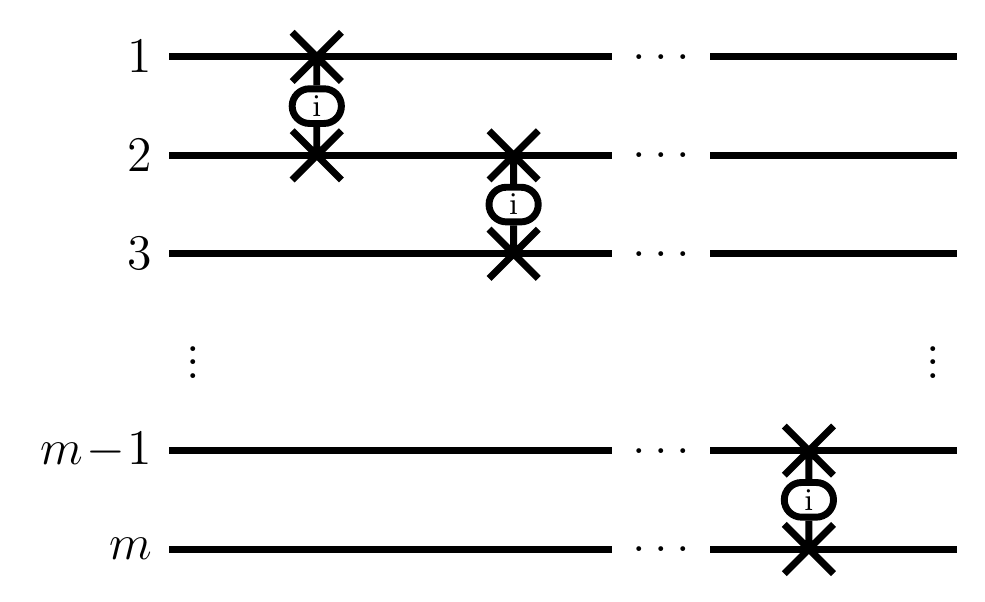}
\caption{%
A chain of \gate{iSWAP} gates in an array of $m$ qubits, as it appears in Fig.~\ref{fig:long-range-CZ-with-iSWAPs}.
}
\label{fig:iSWAP-chain}
\end{figure}

Novel contributions arise when we introduce over-rotations in the \gate{iSWAP}s. An \gate{iSWAP} between qubits $j$ and $k$ can be written as
\begin{align}
\gate{iSWAP}_{j,k} &= \sp j \sm j \sp k \sm k + \sm j \sp j \sm k \sp k \nonumber \\
&\phantom = + \mathrm{i} (\sp j \sm k + \sm j \sp k) \nonumber \\
&= \e^{\mathrm{i} \frac{\pi}{2} (\sp j \sm k + \sm j \sp k)}.
\end{align}
Analogously to the case of \gate{CNOT} chains we consider a chain of \gate{iSWAP}s as shown in Fig.~\ref{fig:iSWAP-chain}. We allow for an over-rotation $\delta \varphi$ to the rightmost gate and, according to the method explained in Sec.~\ref{sec:Method} (Case 2) commute it to the far left. Again, we commute the error gate by gate using Eq.~\eqref{eq:adjoint-commutation}. The first commutation yields
\begin{align}
&\Ad{\gate{iSWAP}_{m-2,m-1}}{\mathrm{i} \delta \varphi (\sp {m-1} \sm m + \sm {m-1} \sp m)} \nonumber \\
={}& \mathrm{i} \big( \sp m (-\sigma^z_{m-1}) \sm {m-2} - \sm m (-\sigma^z_{m-1}) \sp {m-2} \big)\, .
\end{align}
Proceeding iteratively until the error is shifted to the far left, one obtains the effective contribution to the simulated Hamiltonian,
\begin{align}
\delta H (t) &= \frac{n}{\tau} \delta \varphi(t) \, \mathrm{i}^{m-2} \\
&\phantom = \cdot \! \Big(\sp m \prod_{k=2}^{m-1} (-\sigma^z_k) \sm 1 + (-1)^{m-2} \sm m \prod_{k=2}^{m-1} (-\sigma^z_k) \sp 1\Big) \!.
\end{align}

At first glance this looks like a complicated product of many operators, but after transforming it into fermionic language,
\begin{align}
&\sp m \prod_{k=2}^{m-1} (-\sigma^z_k) \sm 1 + (-1)^{m-2} \sm m \prod_{k=2}^{m-1} (-\sigma^z_k) \sp 1 \nonumber \\
={}& c^\dagger_m c^{\phantom \dagger}_1 + (-1)^{m-2} c^\dagger_1 c^{\phantom \dagger}_m,
\end{align}
one finds that the gate errors contribute hopping terms of the form $t_{jk} c^\dagger_j c^{\phantom \dagger}_k$ to the effective Hamiltonian. Analogous results follow in a chain of \gate{-iSWAP} gates when swapping back. This means that in the case of faulty \gate{iSWAP}s we again introduce disorder to the simulated Hamiltonian. While the original Hamiltonian may not include a corresponding term for every pairing $(j, k)$, the additions can then be interpreted as additional transitions. Particularly, unlike in the case of \gate{CNOT} chains, we have a \emph{physical understanding} of the contributions in the effective Hamiltonian due to over-rotations.

\section{Numerical analysis}

\subsection{Verification of the method}
\label{sec:Verification}

The minimal model we have chosen is also suitable for a numerical analysis. For this purpose we choose in the Hamiltonian~\eqref{eq:Hamiltonian-qubits} the following parameters: $U = t_1 = t_2 = g$ and evaluate the time evolution of the initial state $|\psi (t\!=\!0) \rangle = c^\dagger_2 c^\dagger_1 |0\rangle$, where $|0\rangle$ is the vacuum state, up to the time $\tau = 1000/g$. From that we calculate the excitation of the first state $\langle n_1(t) \rangle = \langle c^\dagger_1 (t) c^{\phantom\dagger}_1 (t) \rangle$ and its Fourier transform $\langle n_1(\omega)\rangle$. The time evolution is calculated in several ways:

(1) We use a Trotter expansion according to the algorithm depicted in Fig.~\ref{fig:full-step}, with the $\gate t_2'$ decomposition using \gate{CZ} gates (see Fig.~\ref{fig:t2prime-subst-JW}). We Trotterize the time evolution with a fine step size $\frac{g\tau}{n} = 0.05$. After each Trotter step we evaluate $\langle n_1 (t) \rangle$, such that every step defines a time slice. In each Trotter step we apply random over-rotations to each gate, which are normally distributed with zero mean and different values of the variance.

(2) We use the effective disordered Hamiltonian $H + \delta H$, with $H$ from Eq.~\eqref{eq:Hamiltonian-qubits} and the disorder terms $\delta H(t)$ derived in Sec.~\ref{sec:Algorithm}. We take the same time slices as in (1), and also the time-dependent disorder is chosen in accord with the over-rotations in the respective Trotter steps.

(3) For comparison we also calculate numerically exact the discretized time evolution under the error-free Hamiltonian $H$.

\begin{figure}
\centering
\includegraphics[width=.49\columnwidth]{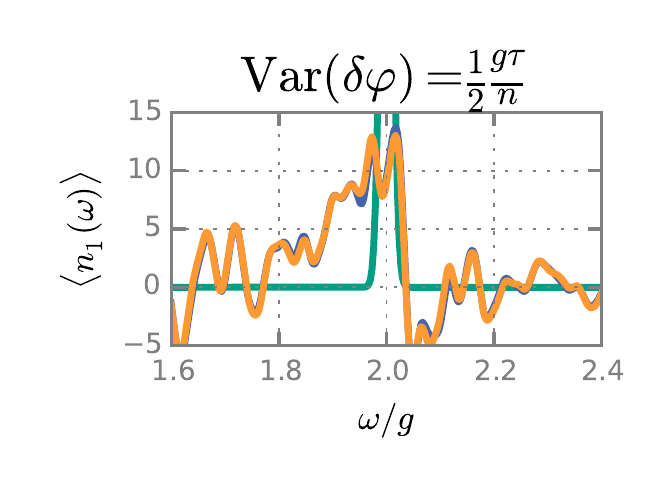}
\includegraphics[width=.49\columnwidth]{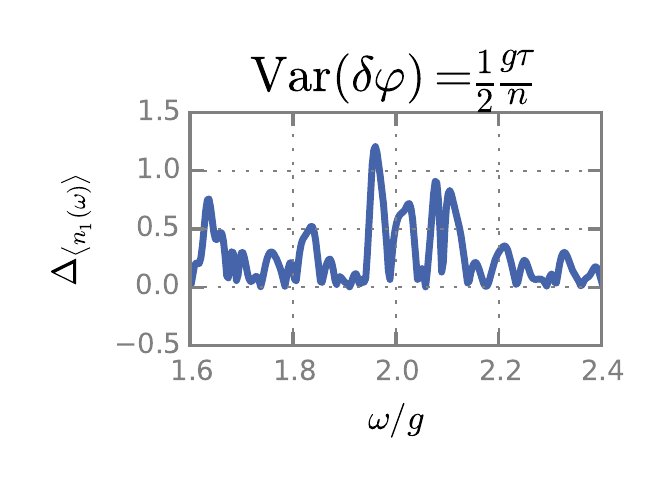}
\caption{%
In the left panel a part of the Fourier spectrum $\langle n_1(\omega) \rangle$ is shown, where the green sharp peak is calculated with the ideal Hamiltonian $H$ (and artificially broadened, see main text). The orange line is the calculation using the algorithm with gate errors by over-rotation, where we chose a large variance $\mathrm{Var}(\delta \varphi) = \frac{1}{2} \frac{g\tau}{n}$ for the statistical gate errors. Under the orange line is a slightly deviating blue line representing the time evolution using the effective Hamiltonian $H + \delta H(t)$. In the right panel the difference $\Delta_{\langle n_1(\omega) \rangle}$ between faulty algorithm and effective Hamiltonian calculation is shown on a much finer scale.
}
\label{fig:plots-single}
\end{figure}

In Fig.~\ref{fig:plots-single} we plot the result for a single run of the time evolution, where the random gate errors have a variance $\mathrm{Var}(\delta \varphi) = \frac{1}{2} \frac{g\tau}{n}$. On the left, a part of the Fourier spectrum $\langle n_1(\omega) \rangle$ is shown. The green sharp peak corresponds to the ideal time evolution without errors. Note that the peak is artificially broadened; $\langle n_1(t) \rangle$ is calculated and multiplied by a Gaussian window function to ensure better convergence of the applied fast Fourier transform. In orange one can see the distorted spectrum calculated using the Trotter expansion with gate errors (according to (1)). It strongly deviates from the ideal result since we have chosen a rather strong disorder. The orange line lies nearly on top of a blue line which represents the result following from the effective disordered Hamiltonian $H + \delta H(t)$. The small difference $\Delta_{\langle n_1(\omega) \rangle}$ between faulty algorithm and effective Hamiltonian with disorder is depicted on the right panel (note the difference in the vertical scale). The comparison shows the validity of the description of the faulty algorithm by an effective disordered Hamiltonian.

\begin{figure}
\centering
\includegraphics[width=.49\columnwidth]{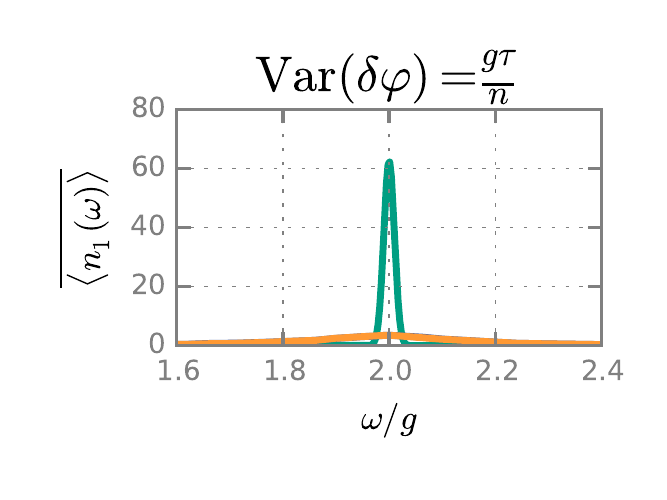}
\includegraphics[width=.49\columnwidth]{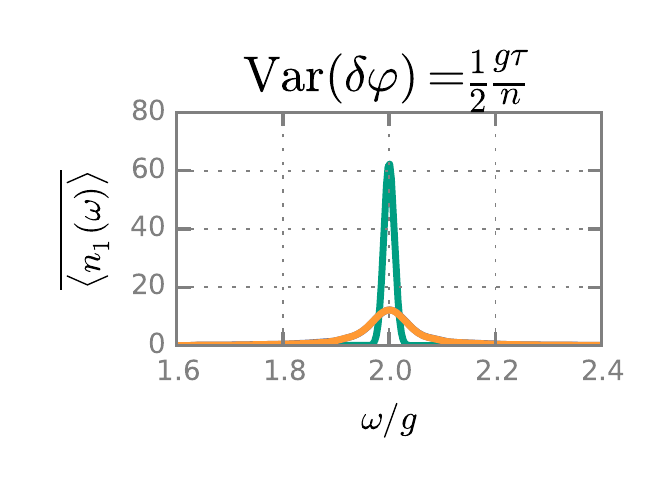}\\
\includegraphics[width=.49\columnwidth]{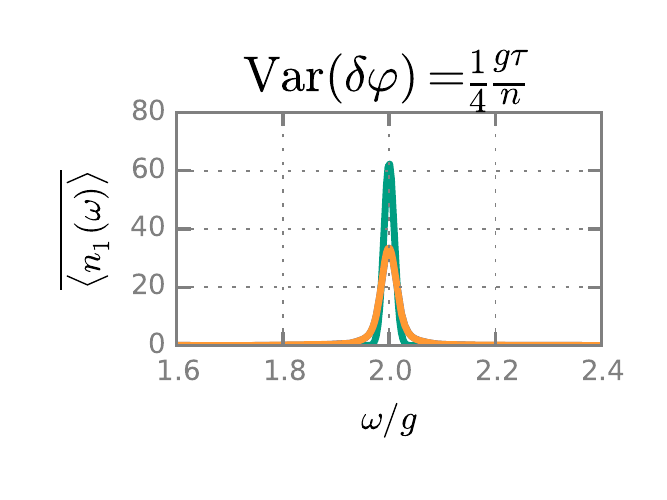}
\includegraphics[width=.49\columnwidth]{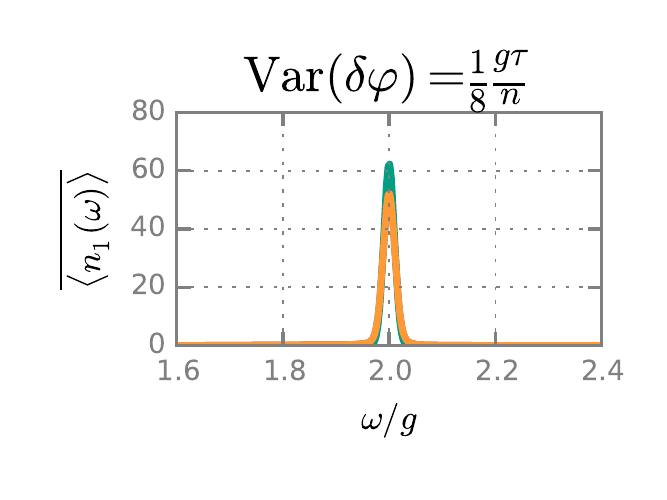}
\caption{%
Shown is the Fourier spectrum $\overline{\langle n_1(\omega) \rangle}$ averaged over many runs of the calculation with different realizations of the random gate errors. The variance of the over-rotations $\mathrm{Var}(\delta \varphi)$ differs in the four plots. The green curves are again the spectrum for the ideal case with artificial broadening (see above), the orange ones show the results of the faulty algorithms. In this case, calculations based on the effective disordered Hamiltonian yield indistinguishable results. We find a broadening of the spectrum which scales with the strength of the over-rotations and corresponds to disorder.
}
\label{fig:plots-avg}
\end{figure}

While for the considered strong gate errors we find a strongly fluctuating distorted result for a single run, averaging over many runs with different random over-rotations produces smoother results. Fig.~\ref{fig:plots-avg} shows the averaged result $\overline{\langle n_1(\omega) \rangle}$ for different variances of the normally distributed errors. In green we depicted again the ideal spectrum (see above), in orange we show the result of the Trotterization with gate errors. After averaging, the description by the effective disordered Hamiltonian and by the Trotterization yield numerically indistinguishable results. As one could expect, we find a broadening of the ideal spectrum, where the broadening grows with the error variance. In the plot with the weakest over-rotations, an order of magnitude lower than the Trotter step size, we find little additional widening of the spectrum.

The numerical analysis showed us that for strong statistical over-rotations in the algorithm the results for individual runs lead to a very distorted spectrum, but averaging over several runs yields a spectrum showing features equivalent to broadening by disorder. Overall we find that gate errors in the algorithm by over-rotations is well described by an effective Hamiltonian with disorder as derived in Sec.~\ref{sec:Method} even for large gate errors.

\subsection{Comparison of different implementations}
\label{sec:Different-implementations}

We continue with the same model system as in the previous subsection and investigate how gate errors in the different implementations of the Jordan-Wigner string discussed in Sec.~\ref{sec:Scaling-up}, using either \gate{CNOT}s or \gate{CZ}s with \gate{iSWAP}s, influence the results. As discussed above we are prepared to discover different physics. We also vary the Trotter step size, as the simulated disorder $\delta H$ grows linearly with the number of Trotter steps (see Eqs.~\eqref{eq:disorder-terms-1},~\eqref{eq:disorder-terms-2}, and Sec.~\ref{sec:Example-algorithm}).

For the comparison we consider the spatial variance of excitations (i.e., charge), $\langle \sigma^2(t) \rangle$, with
\begin{align}
\sigma^2 = \sum_{j=1}^4 r_j^2 \tilde{n}_j - \Big( \sum_{j=1}^4 r_j \tilde{n}_j \Big)^2,
\end{align}
where $\tilde{n}_j = c^\dagger_j c^{\phantom \dagger}_j / (\sum_{k=1}^4 c^\dagger_k c^{\phantom \dagger}_k)$ and we chose $r_1 = 0$, $r_2 = r_4 = 1$, and $r_3 = 2$.\footnote{Note that this choice for $r_i$ represents the minimal number of hoppings to get an excitation from orbital one to orbital $i$. Applied to the minimal Hubbard model (with spin) discussed earlier one might choose the site locations $r_1=r_4=0$, $r_2=r_3=1$. The qualitative conclusions remain unchanged.} This quantity tells us how far excitations in the system are spread out and thus is related to charge diffusion~\cite{steinigeweg_charge_2017,derrico_quantum_2013}.

\begin{figure}
\centering
\includegraphics[width=.49\columnwidth]{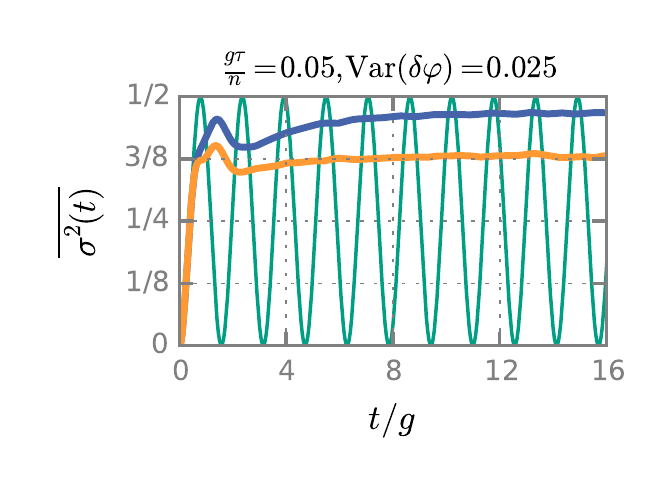}
\includegraphics[width=.49\columnwidth]{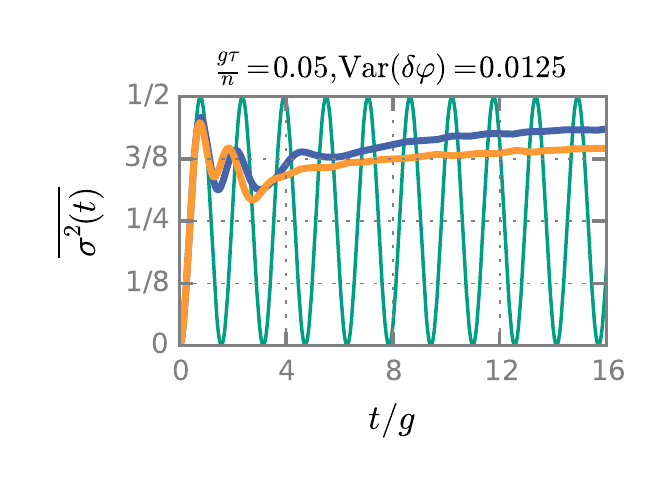}\\
\includegraphics[width=.49\columnwidth]{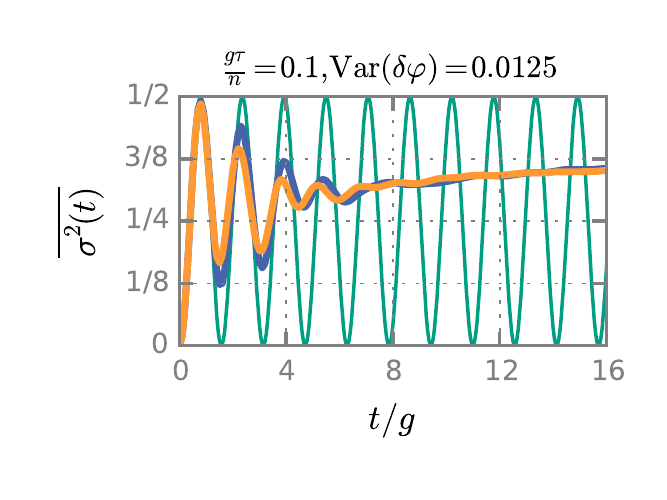}
\includegraphics[width=.49\columnwidth]{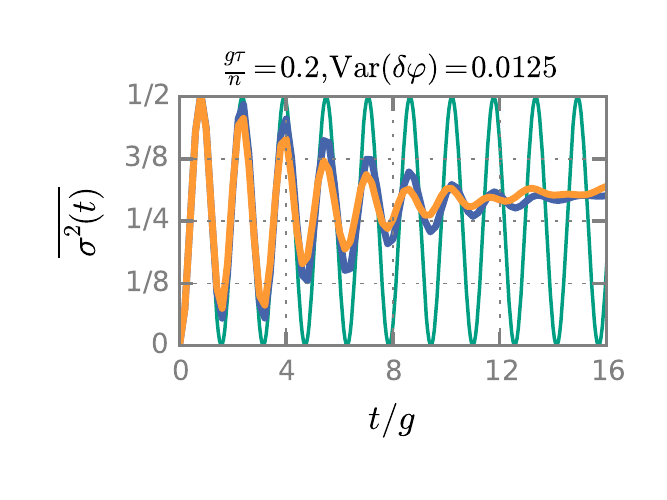}
\caption{
Simulation of the time evolution of the averaged spatial variance $\overline{\langle \sigma^2(t) \rangle}$ (see main text) using different algorithms with random quasi-static gate errors. The Trotter step size $\frac{g\tau}{n}$ and the variance of the over-rotations $\mathrm{Var}(\delta \varphi)$ are altered for each plot. The dashed green line shows the ideal analytic result. The orange line represents a faulty algorithm using \gate{CZ} and \gate{iSWAP} gates for the Jordan-Wigner string. The blue line displays the result with \gate{CNOT} gates for the Jordan-Wigner string. One can see how different algorithms with gate errors result in the simulation of different physics and how the influence of gate errors decreases for larger Trotter step size, i.e., lower gate count.
}
\label{fig:different-implementations}
\end{figure}

The time evolution under the Hamiltonian~\eqref{eq:Hamiltonian-qubits} is simulated with parameters as chosen before $U = t_1 = t_2 = g$, but for the initial state $|\psi(t=0)\rangle = c^\dagger_1 |0\rangle$, i.e., qubit with number 1 is excited and the other ones are in their ground states. We plot $\overline{\langle \sigma^2(t) \rangle}$ over $t/g$, i.e., the averaged result over many runs with random gate errors (normally distributed with zero mean). In this simulation, the errors are chosen to be quasi-static, i.e., for each run a random over-rotation is chosen (for each gate) which does not vary for the Trotter steps, but different runs have different over-rotations. Quasi-static errors are a reasonable noise model for superconducting qubits where the noise spectrum is dominated by low frequencies~\cite{bylander_noise_2011}. The variance of the over-rotations $\mathrm{Var}(\delta \varphi)$ and the Trotter step size $\frac{g\tau}{n}$ are varied. Fig.~\ref{fig:different-implementations} shows the results.

The dashed green line shows the error-free analytic result, where an excitation spreads out through the system but then oscillates back due to the finite system size.

The orange line represents a faulty algorithm using \gate{CZ} and \gate{iSWAP} gates for the Jordan-Wigner string (see Sec.~\ref{sec:Scaling-up}). The induced disorder results in a damping of the oscillation after averaging, i.e., it induces decoherence. The blue line displays the result of a faulty algorithm using \gate{CNOT} gates for modeling the Jordan-Wigner string (see Sec.~\ref{sec:Scaling-up}). As expected from the results of Sec.~\ref{sec:Scaling-up}, in this algorithmic implementation the over-rotations lead to physics which differs from that produced by other algorithms. While we find damping in all cases, the line converges towards a different value.\footnote{Note that the apparent different strength of the damping is due to a different total number of gates in the algorithm. Fewer gates result in fewer disorder terms in the effective Hamiltonian. This discrepancy disappears if one scales for the number of gates accordingly.} This is a direct result of different emerging physics from different kinds of faulty gates; the implementation using \gate{CNOT}s violates particle conservation and the system makes a transition from having a single excitation in it towards half filling. The transition is due to the disorder which randomly excites or relaxes qubits which can be seen as $\sigma^2 = \frac{1}{2}$ for half filling. This illustrates how the choice of algorithm may affect the results of a digital quantum simulation without error correction.

Another aspect illustrated in Fig.~\ref{fig:different-implementations} is the fact that under the influence of gate errors it is -- up to a certain point -- advantageous to reduce the number of Trotter steps and thus of the gate numbers. The disorder of the effectively simulated Hamiltonian scales with their number (see Eqs.~\eqref{eq:disorder-terms-1},~\eqref{eq:disorder-terms-2}, and Sec.~\ref{sec:Example-algorithm}). On the other hand, a coarse Trotterization leads to a large error from the Trotter decomposition. Hence, there is an optimal number of Trotter steps to minimize the simulation error~\cite{Errors_Munro}. One should keep in mind that a finer Trotterization requires higher gate fidelities.

\subsection{Gate fidelity and adiabatic state preparation}
\label{sec:Fidelity-adiabatic-prep}

We conclude this section with two remarks: One about the gate fidelities corresponding to the errors in the above numerical simulations, the other about adiabatic state preparation~\cite{aspuru-guzik_simulated_2005} under the influence of disorder. 

The Trotter step size is opted artificially and the (hardware-dependent) gate error magnitude does not scale accordingly which is why both have to be compared. The error-induced disorder scales linearly with the number of Trotter steps $n$ (see Eqs.~\eqref{eq:disorder-terms-1},~\eqref{eq:disorder-terms-2}, and Sec.~\ref{sec:Example-algorithm}). Note, that this result is independent of the time dependence of the over-rotations $\delta \varphi (t)$, hence, the time dependence of the disorder. Random gate errors with zero mean in each Trotter step influence the simulation, e.g., through damping, less severely than static errors. Regardless of the error type, the gate errors are scaled equally and their strength has to be compared with the Trotter step size.

In our numerical simulations, the variance of the over-rotations $\mathrm{Var}(\delta \varphi)$ was often chosen to be comparable to the Trotter step size $\frac{g\tau}{n}$ and in that sense large. However, since we chose a small step size to ensure the Trotter approximation to be valid, the gate fidelities necessary for this magnitude of error are in fact very high.

The relation between the minimal gate fidelity $\mathcal{F}_\mathrm{min}$ and the strength of over-rotation is easy to compute: As argued in Sec.~\ref{sec:Assumptions}, the application of a gate can be seen as a rotation of the state vector. In the worst case, a state vector orthogonal to the axis of rotation is over-rotated by an angle $\delta \varphi$. In this case, the absolute value of the inner product between ideal state vector and over-rotated state after the application of the gate is $\cos(\delta \varphi)$ (state vectors have unit norm). For a derivation of this see Appx.~\ref{sec:Appx-fidelity}. In the above numerics of Sec.~\ref{sec:Verification} and Sec.~\ref{sec:Different-implementations} we used, e.g., $\mathrm{Var}(\delta \varphi) = 0.025$ and $\mathrm{Var}(\delta \varphi) = 0.0125$, which corresponds to a averaged minimal gate fidelity $\overline{\mathcal{F}}_\mathrm{min}$ of $99.969\,\%$ and $99.992\,\%$, better than the values reached by present-day devices~\cite{Marinis_Threshold,Blatt_error_correction}.

Let us investigate the situation for a gate fidelity of $99.0\,\%$. This corresponds to a variance of over-rotations of $\mathrm{Var}(\delta \varphi) = \arccos(0.99) \approx 0.142$. Looking at Eqs.~\eqref{eq:disorder-terms-1},~\eqref{eq:disorder-terms-2}, and Sec.~\ref{sec:Example-algorithm} one finds that the disorder terms are scaled with $\frac{n}{\tau}$. Hence, to compare the strength of the disorder with the energy scale $g$ of the Hamiltonian, one has to compare the over-rotation variance $\mathrm{Var}(\delta \varphi)$ with the Trotter step size $\frac{g\tau}{n}$.

There is a natural time scale of the dynamics of the Hamiltonian associated with its energy scale $g$. To resolve these dynamics one has to simulate times up to $\tau \sim 1/g$. This yields a Trotter step size $\frac{g\tau}{n} \sim \frac{1}{n}$. For a gate fidelity of $99.0\,\%$, which means $\mathrm{Var}(\delta \varphi) \approx 0.142$, this value quickly becomes comparable to (or larger than) $\frac{1}{n}$ for increasing $n$. Accordingly, the number of Trotter steps is very limited for current experimental gate fidelities~\cite{Errors_Munro}. In fact, for the algorithm of the previous subsection (Sec.~\ref{sec:Different-implementations}) we found that a gate fidelity above $99.99\,\%$ is necessary to allow for enough Trotter steps while not producing an overdamped result.

\begin{figure}
\centering
\includegraphics[width=.49\columnwidth]{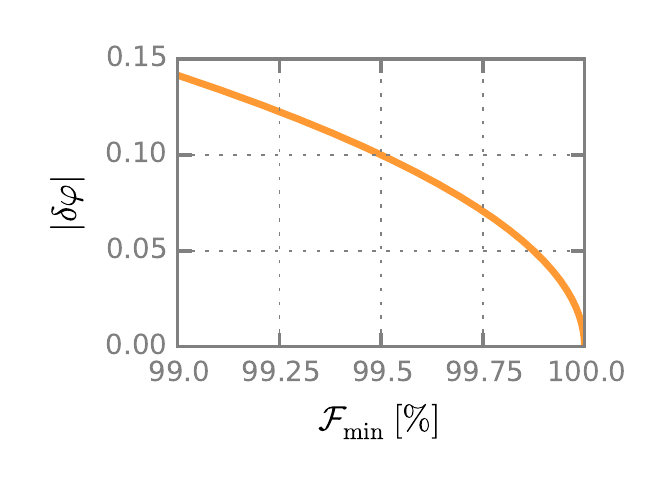}
\includegraphics[width=.49\columnwidth]{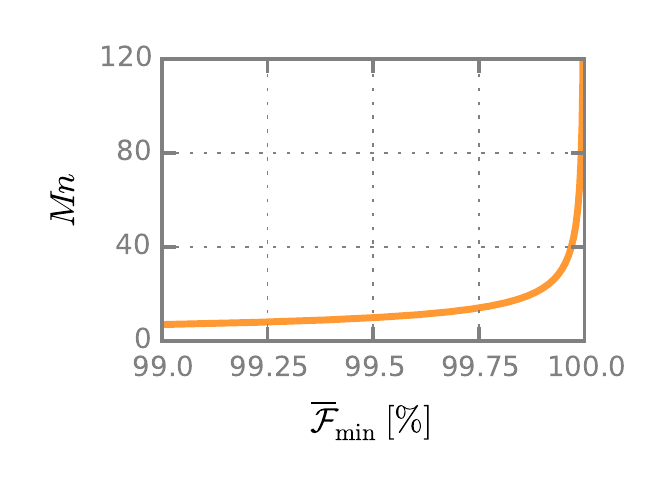}
\caption{
On the left is the over-rotation magnitude $|\delta \varphi|$ plotted over the minimal gate fidelity $\mathcal{F}_\mathrm{min}$ of an over-rotated gate. The steep slope of $|\delta \varphi| = \arccos(\mathcal{F}_\mathrm{min}) \approx \sqrt{2(1-\mathcal{F}_\mathrm{min})}$ as $\mathcal{F}_\mathrm{min}$ approaches one requires gate fidelities to be very high for the disorder in the effectively simulated Hamiltonian to be sufficiently low (the disorder strength scales linearly with $|\delta \varphi|$). The right plot shows the total number of faulty gates $Mn$ that can be run for a given average minimal gate fidelity $\overline{\mathcal{F}}_\mathrm{min}$ according to Eq.~\eqref{eq:total-number-gates}. $M$ is the number of faulty gates per Trotter step and $n$ the number of Trotter steps. Using this rough estimate, one can see how fidelities have to be very high in order to allow for a large number of gates.
}
\label{fig:arccosplot}
\end{figure}

As explained in Appx.~\ref{sec:Appx-fidelity}, the magnitude of over-rotation can be expressed as $|\delta \varphi| = \arccos(\mathcal{F}_\mathrm{min})$. The absolute value $|\delta \varphi|$ is in fact a metric called the Bures metric. In our error model it is a much more useful measure for gate errors, since it linearly contributes to the strength of disorder in the effectively simulated Hamiltonian, and can be compared meaningfully with the Trotter step size $\frac{g\tau}{n}$. The minimal gate fidelity $\mathcal{F}_\mathrm{min}$ is an unintuitive measure, as illustrated in Fig.~\ref{fig:arccosplot}: It holds that $|\delta \varphi| = \arccos(\mathcal{F}_\mathrm{min}) \approx \sqrt{2(1-\mathcal{F}_\mathrm{min})}$ for $\mathcal{F}_\mathrm{min} \approx 1$. The diverging slope of the square root requires $\mathcal{F}_\mathrm{min}$ to get increasingly close to one to further reduce the disorder strength of the effective Hamiltonian.

Averaging over a normally distributed $\delta \varphi$ with zero mean we find
\begin{align}\label{eq:over-rotation-variance-and-fidelity}
\mathrm{Var}(\delta \varphi) = \sqrt{2(1-\overline{\mathcal{F}}_\mathrm{min})},
\end{align}
where we denote $\overline{\mathcal{F}}_\mathrm{min}$ the averaged minimal gate fidelity. With this we can derive an estimate for the maximum number of Trotter steps for a given algorithm and gate fidelity: As mentioned above as well as in Eqs.~\eqref{eq:disorder-terms-1},~\eqref{eq:disorder-terms-2}, and Sec.~\ref{sec:Example-algorithm}, the disorder terms in the Hamiltonian are of magnitude $\frac{n}{\tau} \mathrm{Var}(\delta \varphi)$. There are a total of $M$ such terms, i.e., $M$ faulty (two-qubit) gates per Trotter step. The total disorder is therefore of magnitude $\frac{Mn}{\tau} \mathrm{Var}(\delta \varphi)$.\footnote{Note that this estimate involves the triangle inequality $|| \sum_{j=1}^M A_j || \leq \sum_{j=1}^M || A_j || = M$ (here $||A_j|| = 1$ for all $j$). It is therefore a worst-case scenario where all gate errors of independent gates add up adversely. The number $M$ could effectively be much lower.} The disorder strength should be (much) lower than the Hamiltonian energy scale $g$. If we again set $\tau \sim \frac{1}{g}$, and use Eq.~\eqref{eq:over-rotation-variance-and-fidelity} for the over-rotation variance we arrive at
\begin{align}\label{eq:total-number-gates}
Mn < \frac{1}{\sqrt{2(1-\overline{\mathcal{F}}_\mathrm{min})}}.
\end{align}
This yields an estimate for the maximum number of Trotter steps $n$, resp.\ the maximum total number of faulty (two-qubit) gates in an algorithm $Mn$. It is a rough, worst-case estimate but suggests that gate fidelities should be well above $99.9\,\%$ to allow for a larger number of gates (see Fig.~\ref{fig:arccosplot}).

With the relation between gate fidelity and over-rotations at hand, we also studied how adiabatic state preparation would perform with current state-of-the-art fidelities. Unfortunately, we found the time-dependent disorder induced by the gate errors to be very problematic during an adiabatic time evolution. During the adiabatic evolution the system is required to stay in an eigenstate while the system parameters are slowly changing (e.g., the interaction $U$ is adiabatically turned on). Gaps in the spectrum protect from crossings to a different eigenstates as long as the time scale of the evolution is long enough. If the time evolution is influenced by disorder, crossings may occur if the energy scale of the disorder is comparable to the gap size in the spectrum.

In numeric calculations we found that even very low disorder prohibits an adiabatic evolution to a final state with the correct properties. Due to the large number $n$ of Trotter steps in a (slow) adiabatic time evolution, increasingly high fidelities would be needed to keep the disorder negligibly low, since it grows linearly with $n$ (see Eqs.~\eqref{eq:disorder-terms-1},~\eqref{eq:disorder-terms-2}, and Sec.~\ref{sec:Example-algorithm}).

Hence, we would suggest different approaches for the state preparation in a digital quantum simulation without error correction, e.g., variational approaches~\cite{wecker_progress_2015,mcclean_theory_2016}. These algorithms need significantly less gates than an adiabatic time evolution. The much lower number of Trotter steps required reduces the disorder introduced in the simulation. Furthermore, these approaches can generally mitigate gate errors~\cite{colless_computation_2018}. It appears promising that variational methods could allow for efficient ground state preparation even with faulty gates.

\section{Conclusion}

In this work we analyzed the effect of gate errors on a digital quantum simulation, when the time evolution of a system with Hamiltonian $H$ is evaluated using algorithms based on the Trotter expansion. We showed that gate errors due to over-rotations effectively introduce an extra term $\delta H$ in the Hamiltonian, which in many cases can be interpreted as a disorder term. However, we also demonstrated that the nature of these contributions depends on the choice of the algorithms and that different algorithms may introduce different physics in a faulty simulation.

The method was then applied to an example system, a Fermi-Hubbard model with spin-flip interaction. The simulation of this model was translated into an algorithm based on various two-qubit gates. We showed that due to gate errors effectively a disordered fermionic system is simulated. This helps interpreting the results of quantum simulations without quantum error correction. The example also demonstrates how the effects of over-rotations depend on the choice of algorithms. For instance in the example of Sec.~\ref{sec:Algorithm} we showed that replacing \gate{CZ} gates by a chain of \gate{CNOT}s introduces unphysical multi-particle interactions violating particle conservation.

We also illustrated our findings by a numerical analysis of a small model system.
We stress that the method can be extended beyond the small system. In such cases the algorithms can be performed and analyzed piecewise. In particular, the findings of Sec.~\ref{sec:Algorithm} are valid for a large class of Hubbard-like Hamiltonians with general hopping terms $t_{jk} c^\dagger_j c^{\phantom \dagger}_k$ and density-density interactions $V_{jk} c^\dagger_j c^{\phantom \dagger}_j c^\dagger_k c^{\phantom \dagger}_k$. This covers a wide variety of problems in quantum chemistry and solid state physics.

The method can also be extended to higher order Trotter expansions. For the second order expansion one has to require that over-rotations are restricted to $|\delta \varphi| \leq (\frac{g\tau}{n})^2$.

We established a connection between the over-rotation of gates and their fidelities in Sec.~\ref{sec:Fidelity-adiabatic-prep}. This enables us to quantify the strength of the disorder in the effective Hamiltonian for a given fidelity and Trotter step size. We have to conclude that fidelities as presently achieved in experiments impose severe limitations. The error rates should decrease by orders of magnitude to allow for a widely useful number of Trotter steps, without introducing too much disorder in the simulation. The induced disorder is particularly harmful to adiabatic state preparation.

Following Eq.~\eqref{eq:total-number-gates}, we find that gate fidelities of $99\,\%$ only allow for something like $10$ gates to run in an algorithm. To reach of the order of $100$ gates, fidelities should improve towards $99.99\,\%$. We would like to stress again at this point that our modelling provides a worst-case estimate, and the possible maximum gate count could in fact be higher. For instance, the effective number of faulty gates $M$ per Trotter step in Eq.~\eqref{eq:total-number-gates} may be smaller than the number of two-bit gates. In addition, our estimates are based on the assumption that gate errors are solely caused by over-rotations. In reality, there is a variety of error sources that contribute to lowering the measured gate fidelities. Other sources of errors and a combination of different errors might lead to a different scaling behavior, especially when gates can, e.g., be run in parallel. While this does not make any difference in the case of over-rotations, it could greatly reduce the impact of decoherence. In addition, variational algorithms are promising for short term application due to their low number of necessary Trotter steps and their potential to mitigate errors.

Overall, we demonstrated that the effects of gate errors in a digital quantum simulation of (fermionic) systems can be understood on the level of modified Hamiltonians, and the strength of the contributions from errors can be quantified for a given Trotter step size and gate fidelity.

\begin{acknowledgments}
This work was supported by the DFG Research Grant SCHO 287/7-1.
\end{acknowledgments}

\section*{Appendix}

\begin{appendix}

\section{Fidelity of over-rotated gates}
\label{sec:Appx-fidelity}

In this appendix, we derive the relation between the minimal gate fidelity $\mathcal{F}_\mathrm{min}$ of faulty gate due to an over-rotation and the angle of over-rotation $\delta \varphi$. 
The gate is defined by the unitary operator
\begin{align}
U(\varphi) = \e^{\mathrm{i} \varphi A},
\end{align}
where $A$ is a Hermitian operator with a finite spectrum consisting of the eigenvalues $\lambda_1,\dots,\lambda_d$, where $|\lambda_j| \leq 1$, $\lambda_1 = 1$, and $\lambda_n = -1$. Note that this is the case if $A$ is a product of Pauli operators. In general, every finite Hermitian operator can be cast into this form through rescaling and shifting by constants. The faulty gate is represented by $U(\varphi + \delta \varphi)$, with an over-rotation angle $\delta \varphi$. Also the faulty gate error is therefore a unitary transformation.

With a unitary error model it is sufficient to regard pure states as input states. The fidelity between two pure states $|\psi\rangle$ and $|\phi\rangle$ is given by their overlap $| \langle \psi | \phi \rangle |$. Given an input state $|\psi\rangle$, the application of the gate would ideally yield the final state $U(\varphi) |\psi\rangle$ whereas the application of the over-rotated gate yields $U(\varphi + \delta \varphi) |\psi\rangle$. The minimal gate fidelity $\mathcal{F}_\mathrm{min}$ is given by the overlap of the ideal and faulty final state, minimized over all input states, hence:
\begin{align}
\mathcal{F}_\mathrm{min} &= \min_{|\psi\rangle} |\langle \psi | U^\dagger (\varphi) U(\varphi + \delta \varphi) | \psi \rangle |\nonumber \\
&= \min_{|\psi\rangle} |\langle \psi |\e^{\mathrm{i} \delta \varphi A} | \psi \rangle |.
\end{align}

We investigate the quantity to be minimized: One can diagonalize the Hermitian matrix $A = \mathrm{diag}(\lambda_1,\dots,\lambda_d)$ and find its eigenbasis $\{|\lambda_1\rangle,\dots,|\lambda_d\rangle\}$. Writing $|\psi\rangle = \sum_{j=1}^d a_j |\lambda_j\rangle$ with $\sum_{j=1}^d |a_n|^2 = 1$ results in
\begin{align}
|\langle \psi | e^{\mathrm{i} \delta \varphi A} | \psi \rangle | = |\sum_{j=1}^d |a_j|^2 e^{\mathrm{i} \delta \varphi \lambda_j}|.
\end{align}
Making use of the fact that the absolute value of a complex number is smaller than the magnitude of its real part for the expression $e^{\mathrm{i} \delta \varphi \lambda_j} = \cos(\delta \varphi \lambda_j) + \mathrm{i} \sin(\delta \varphi \lambda_j)$, we arrive at
\begin{align}
|\langle \psi | e^{\mathrm{i} \delta \varphi A} | \psi \rangle | \geq |\sum_{j=1}^d |a_j|^2 \cos(\delta \varphi \lambda_j)|.
\end{align}
It is reasonable to restrict $|\delta \varphi| \leq \pi/2$, since an over-rotation as a gate error should be small. As $|\lambda_j| \leq 1$ it follows that $\cos(\delta \varphi \lambda_j) \geq \cos(\delta \varphi) \geq 0$, therefore
\begin{align}
|\langle \psi | e^{\mathrm{i} \delta \varphi A} | \psi \rangle | \geq \cos(\delta \varphi) \sum_{j=1}^d |a_j|^2 = \cos(\delta \varphi).
\end{align}

This establishes a lower bound of $\mathcal{F}_\mathrm{min}\ge \cos(\delta\varphi)$. For a special state
\begin{align}
|\psi_\mathrm{min}\rangle = \frac{1}{\sqrt{2}} (|\lambda_1\rangle + |\lambda_d \rangle)
\end{align}
we indeed find
\begin{align}
|\langle \psi_\mathrm{min} | e^{\mathrm{i} \delta \varphi A} | \psi_\mathrm{min} \rangle | = \cos(\delta \varphi),
\end{align}
since $\lambda_1 = 1$ and $\lambda_d =-1$. Hence, the lower bound is actually tight and we obtain
\begin{align}\label{eq:minimal-fidelity}
\mathcal{F}_\mathrm{min} = \cos(\delta \varphi).
\end{align}

This concludes our derivation. Note, that the over-rotation magnitude $|\delta \varphi| = \arccos(\mathcal{F}_\mathrm{min})$ is in fact a metric, namely the Bures metric (resp.~the Fubini-Study metric).

In an experiment $\delta \varphi$ is not fixed but rather a statistical variable. In this case it is sensible to consider the average minimal gate fidelity $\overline{\mathcal{F}}_\mathrm{min}$, i.e., the expectation value of $\mathcal{F}_\mathrm{min}$ for a given distribution of $\delta \varphi$. For small $\delta \varphi$ one can expand Eq.~\eqref{eq:minimal-fidelity}, $\mathcal{F}_\mathrm{min} \approx 1 - \frac{\delta \varphi^2}{2}$, such that for a normally distributed $\delta \varphi$ with zero mean and variance $\mathrm{Var}(\delta \varphi)$ we find $\overline{\mathcal{F}}_\mathrm{min} = 1 - \frac{\mathrm{Var}(\delta \varphi)^2}{2}$.

\end{appendix}

%

\end{document}